\documentclass[10pt,letterpaper]{article}
\usepackage[top=0.85in,left=2.75in,footskip=0.75in]{geometry}

\usepackage{amsmath,amssymb}

\usepackage{changepage}

\usepackage{textcomp,marvosym}

\usepackage{cite}

\usepackage{nameref,hyperref}

\usepackage[right]{lineno}

\usepackage[nopatch=eqnum]{microtype}
\DisableLigatures[f]{encoding = *, family = * }

\usepackage[table]{xcolor}

\usepackage{array}

\newcolumntype{+}{!{\vrule width 2pt}}

\newlength\savedwidth

\usepackage{setspace} 
\raggedright
\usepackage[aboveskip=1pt,labelfont=bf,labelsep=period,justification=raggedright,singlelinecheck=off]{caption}

\makeatletter
\renewcommand{\@biblabel}[1]{\quad#1.}
\makeatother

\usepackage{lastpage,fancyhdr,graphicx}
\usepackage{epstopdf}
\fancyheadoffset[L]{2.25in}
\fancyfootoffset[L]{2.25in}
\usepackage{enumitem} 
\usepackage{dsfont}
\usepackage[prependcaption,colorinlistoftodos]{todonotes}
\usepackage{adjustbox}
\usepackage{amsthm}
\usepackage{subcaption}
\usepackage{diagbox}

\newcommand{\vsa}{\vspace*{-0.40cm}}
\newcommand{\vsb}{\vspace*{-0.20cm}}

\newcommand{\vect}[1]{\mathds{#1}}

\newcommand{\ones}[1][]{\vect{1}_{#1}}

\newcommand{\real}{\mathbb{R}}
\newcommand{\realnonnegative}{\mathbb{R}_{\geq0}}
\newcommand{\matrixSym}[1]{\uppercase{#1}}
\newcommand{\vleft}{v_{\mathrm{L}}}

\newcommand{\eqrefreg}[1]{\textup{{\normalfont[\ref{#1}}\normalfont]}}
\renewcommand{\eqref}[1]{\textup{{\normalfont Eq.~[\ref{#1}}\normalfont]}}

\newcommand{\setdef}[2]{\{#1\; | \; #2\}}

\newcommand{\map}[3]{#1: #2 \rightarrow #3}

\DeclareMathOperator{\diag}{diag}
\newtheorem{theorem}{Theorem}
\newtheorem{lemma}[theorem]{Lemma}

\newtheorem{hyp}{Hypothesis}

\usepackage{etoolbox}
\usepackage{booktabs} 

\AtBeginEnvironment{tabular}{
    \sffamily\fontsize{8}{10}\selectfont
}

\begin{document}
\vspace*{0.2in}

\begin{flushleft}
{\Large
\textbf\newline{Expertise and confidence explain how social influence evolves along intellective tasks} 
} 
\newline
\\
Omid Askarisichani\textsuperscript{a,*},
Elizabeth Y. Huang\textsuperscript{b,c},
Abed K. Musaffar\textsuperscript{b,c},
Noah E. Friedkin\textsuperscript{c,d},
Francesco Bullo\textsuperscript{b,c},
Ambuj K. Singh\textsuperscript{a}
\\
\bigskip
\textbf{a} Department of Computer Science, University of California, Santa Barbara, CA 93106
\\
\textbf{b} Department of Mechanical Engineering, University of California, Santa Barbara, CA 93106
\\
\textbf{c} Center for Control, Dynamical Systems, and Computation, University of California, Santa Barbara, CA 93106
\\
\textbf{d} Department of Sociology, University of California, Santa Barbara, CA 93106
\bigskip

%
%





* bullo@ucsb.edu
\end{flushleft}


\section*{Abstract}
Discovering the antecedents of individuals' influence in collaborative environments is an important, practical, and challenging problem.  In this paper, we study interpersonal influence in small groups of individuals engaged in a series of cognitive tasks.  We observe that along an issue sequence with feedback, individuals with higher expertise and social confidence are accorded higher interpersonal influence.  Additionally, we identify a tendency among underperforming individuals to underestimate the expertise of their high-performing peers.

Building upon these insights, we put forth three hypotheses and substantiate their validity through a blend of empirical evidence and theoretical reasoning.  We report empirical evidence on longstanding theories of transactive memory systems, social comparison, and confidence heuristics on the origins of social influence.  To articulate the dynamic nature of how individuals fine-tune their interpersonal influence over time, we introduce a cognitive dynamical model inspired by these theories. We demonstrate the model's accuracy in predicting individuals' influence and provide analytical results on its asymptotic behavior for identically performing individuals.

Lastly, we propose a novel approach for predicting the influence of individuals, based upon deep neural networks in conjunction with a pre-trained text embedding model.  By analyzing message contents, message timings, and individual accuracy during tasks, we are able to accurately predict individuals' self-reported influence over time.  Extensive experiments verify the accuracy of the proposed models compared to benchmarks such as structural balance and reflected appraisal models. While the neural networks model boasts the highest predictive accuracy, the dynamical model stands out for its interpretability in the context of influence prediction.

\section*{Introduction}
Inevitably, relationships among collaborating actors evolve over time, with
people changing their opinions or appraisals of one another. Such
relationships form a network structure called an influence/appraisal
network whose edges may portray trust/distrust, friendship/enmity, and
like/dislike~\cite{MHDG:74, FJ90-SIO, das2014modeling, NEF-FB:16d,
  jia2015opinion, ravazzi2017learning, zheng2015social, OA-FB-NEF-AKS:22}. Investigations of
the evolution of such networks draw on a rich body of literature on opinion
dynamics~\cite{degroot1974reaching,FJ90-SIO,altafini2012consensus,proskurnikov2017tutorial,DC:22}. However,
these opinion dynamics models assume that the influence network of a group
is given {\em a priori}. Our goal is to quantitatively estimate the
influence network among individuals along a sequence of issues.  As the
influence network among the individuals is represented by a row-stochastic
matrix, the estimation of this matrix paves the way for solving problems
such as influence maximization~\cite{chen2009efficient, apc2015}; viral
marketing~\cite{leskovec2007dynamics, chen2010scalable}; personalized
recommendation~\cite{song2006personalized,YW-JK:23}; feed
rankings~\cite{ienco2010meme}; target advertisement~\cite{li2015real}; and
selecting influential tweeters and blogs~\cite{weng2010twitterrank,
  bakshy2011everyone, leskovec2007cost}.
Classic studies of the antecedents of interpersonal influence include French and Raven's work~\cite{JRPF-BR:59} on the bases of social power, and cognitive biases research~\cite{goldsmith2015social} showed that individuals are accorded influence based on their job titles, past performance, friends' opinions, etc. There has also been mathematical modeling of the endogenous evolution of appraisal networks. Friedkin \emph{et al.}~\cite{NEF:11} showed how reflected appraisal mechanisms elevate or dampen the self-weights of group members along a sequence of issues. Jia \emph{et al.}~\cite{jia2015opinion} proposed the DeGroot-Friedkin model, where the appraisal network evolves as a function of the social power within the group. Jia \emph{et al.}~\cite{jia2016coevolution} also studied how over time, the coevolution of appraisal and influence networks leads to a generalized model of structural balance theory~\cite{heider1946attitudes, marvel2011, srinivasan2011local, facchetti2011computing, szell2010, nc2019}. Mei \emph{et al.}~\cite{mei2017dynamic} modeled collective learning in teams of individuals using appraisal networks, where the appraisal dynamics change as a function of the performance of individuals within the team.

Research on transactive memory systems (TMS)~\cite{austin2003transactive, lewis2003measuring, lewis2004knowledge, lewis2005transactive,GDP-EM:22} provides an approach to the formation of influence systems. As members observe the  performances of each other, their understanding of ``who knows what'' tends to converge to an accurate assessment, leading to greater coordination and integration of members’ skills~\cite{yuan2010impact, lewis2004knowledge, liang1995group}. Empirical research~\cite{austin2003transactive, lewis2003measuring, lewis2004knowledge, lewis2005transactive} across a range of team types and settings demonstrates a strong positive relationship between the development of a team TMS and team performance~\cite{laughlin1986demonstrability, laughlin1980social,YT-LW:23}.


Research in social comparison theory has shown that individuals tend to evaluate their own abilities by biased comparisons with their peers. In particular, Woods~\cite{wood1989socialcomparisontheory} describes several motivations behind biased social comparison such as self-esteem protection~\cite{shepperd1993productivity}, lack of appropriate incentives~\cite{shepperd1989individual}, or the existence of dominant individuals who skew member contributions~\cite{beatty1996using, davison2014individual}. Davison \emph{et al.}~\cite{davison2014individual}'s experimental results show that low-performing individuals tend to overestimate (resp. underestimate) low-performers (resp. high-performers), i.e. high-performing individuals are better able to recognize other experts than low-performing individuals. To study such psychological cognitive biases, several scientists have conducted group experimental studies by self and peer evaluations~\cite{beatty1996using, drexler2001peer, miller2000self}.

Research on confidence heuristics~\cite{thomas1995confidence, price2004intuitive} has shown that the more self-confident individuals are, the more influence they are accorded by others. London \emph{et al.}~\cite{london1970jury} find that ``the single significant behavioral difference between persuaders and persuadees was in the expression of confidence''. Confidence heuristics is defined based on a social and psychological norm, whereby more confidently expressed arguments signal better information, allowing an efficient revelation of information and decision-making based on expressed confidence~\cite{thomas1995confidence}.

We build on the above three lines of research on influence networks, TMS, and biases/heuristics. Although the problem of estimating social power~\cite{jia2015opinion, tnse2017} and influence networks have been studied before~\cite{castro2018particle, castro2018influence, smith2018influence}, existing research lacks empirical studies as they are mostly based on theory and grounded only on simulation-based analyses~\cite{mei2017dynamic}. Moreover, previous studies on influence estimation have focused on proxies of influence such as propagation of hashtags, quotes, and retweets~\cite{smith2018influence, gomez2016influence, du2013scalable, deng2012your, ye2010measuring, riquelme2016measuring}. A recent influential study by Almaatouq \emph{et al.}~\cite{almaatouq2020adaptive} found that social influence is significantly correlated with confidence and correctness. However, no estimation method is proposed that mathematically formulates how these factors contribute to the underlying dynamics of influence. Furthermore, we find these two factors alone do not lead to the most accurate predictions of self-reported influence. 

Studies on empirical estimation of the weighted network of who-influences-whom are rare~\cite{manteli2014effect, huang2018exploring}. In the present work, we probe more deeply into the foundations of the links between individual performance, self-confidence, and social comparison on interpersonal influence. Our work bridges the gap between empirical and simulation-based results by utilizing sociology-inspired mechanisms and machine-learning based models to estimate social influence in groups. Overall, to the best of our knowledge, this study is the first to estimate influence matrices through conversations in human subject experiments, where teammates cooperate to solve intellective tasks.
Our contributions in this paper are threefold.
First, we find empirical support for widely established theories in psychology, sociology and management regarding the effect of TMS~\cite{austin2003transactive, lewis2003measuring, lewis2005transactive}, confidence heuristics~\cite{thomas1995confidence, price2004intuitive} and social comparison theory~\cite{davison2014individual, wood1989socialcomparisontheory} on individuals' influence over their teammates.
Second, we introduce a novel cognitive dynamical model based on the aforementioned theory regarding how influence is accorded. This cognitive model is validated against the empirical data and can be used to estimate influence matrices in sequential experimental design. We provide simulation results on the asymptotic behavior of the model and analytical results for the case with identically performing individuals.

Third, we propose a machine learning-based model for estimating influence networks using features such as individual performance, past influence matrices, and communication contents as well as communication timestamps. Extensive experiments show that our proposed neural network model surpasses all baseline methods.


\section*{Materials and methods}
\subsection*{Ethics Statement}
The research protocol and consent procedures were approved by the Institutional Review Board of the University of California, Santa Barbara (IRB \#21-18-0973 titled ``Quanta: Quantitative Network-Based Models of Adaptive Team Behavior'') as well as by Army Research Lab Human Research Protection Official (ARL \#18-154). All human subjects provided informed written consent prior to study participation. Recruitment for this study began on July 01, 2019 and the last experiment concluded on June 03, 2019.

\subsection*{Experimental design}

We collected data for 31 teams comprising of four human members each. Each team is presented with the same sequence of 45 truth questions that fall into three categories: Science and Technology, History and Mythology, and Literature and Media. Every team has two minutes to answer each question. First, team members answer individually before the answers are revealed to the team. Second, they are asked to collaborate on a single unanimous response. Lastly, the platform reveals the correct answer immediately after a team submits their answer. Thus, they are provided with immediate feedback on their performance after every response. The design also incorporates a multi-part incentive for subjects to seek the correct answer on each question: an evolving team performance score; an option to consult with one of four available AI-agents after the team discussion (the AI-agents may or may not provide a correct answer) which, if exercised, must lower the team's performance score regardless of whether the agent provides a correct or incorrect answer; and feedback to each team on correct and incorrect answers. This multi-part incentive structure is operated to concentrate the attention of the team on the evaluation of the relative expertise of its members.


In this study, each experiment consists of nine rounds of five intellective questions each, with teams surveyed after each round. At the end of each round, subjects are asked to record the influence of their teammates in their decision-making process as a percent value, such that the sum of all values adds up to 100. Thus, the number of chips that a subject allocates to a particular member should indicate the relative extent to which they were personally influenced by that member. The number of chips that subjects allocate to themselves should indicate the extent to which their final answer was not affected by the conversation.


After normalization, the self-reported interpersonal influences form a row-stochastic influence matrix for every round, containing only non-negative entries (every row sums up to one in a row-stochastic matrix). The platform ensures that in each inquiry, the reported influence matrix has non-negative entries and is row-stochastic.  Additionally, the platform collects the instant messages including time and content of every message, the individual and group answers, and the self-reported influence matrices.

Since the platform displays the correct answer to every question immediately after the group submits its response, attentive subjects can use the individual responses and text discussion to keep track of which individual teammates may have expertise in one or more areas over the course of the experiment. Thus, along the problem sequence, the team may solve problems more efficiently and also more accurately.

\subsection*{Models for estimating interpersonal influence matrix}
In this section, we provide mathematical formulation regarding how the proposed dynamical model, linear, and deep neural network model alongside baselines efficiently frame and solve the influence matrix estimation problem. The proofs for all lemmas are provided in the \nameref{S1_Appendix}.

\subsubsection*{Proposed cognitive dynamical models}
We propose various discrete time dynamical models that characterize the evolution of the influence network based on various sociological concepts. Let the simplex be defined as $\Delta_n = \{x\in\realnonnegative \,|\, \ones[n]^\top x=1\}$. Given an estimate of a previous row-stochastic influence matrix $\hat M^{(t)}$  and an estimate of normalized or perceived expertise $x^{(t)}\in\Delta_n$, our models are in the general form
\begin{equation*}
    \hat M^{(t+1)} = T\big(\hat M^{(t)},x^{(t)}\big) \quad\text{for } t\geq 1,
\end{equation*}
\noindent where $\hat M^{(t)}$ denote the influence matrix estimate at round $t$. 
Let $\hat M_d^{(t)} = [\hat M_{11}^{(t)},\dots, \hat M_{nn}^{(t)}]^\top$ denote the vector of self-influence weights.
Consider the normalized expertise and perceived expertise, defined as follows. \emph{Normalized expertise} $\bar y^{(t)} \in \Delta_n$ is defined as 
\begin{equation}
\label{eq:expertise-normalized}
    \bar y^{(t)} = \big(\ones[n]^\top y^{(t)}\big)^{-1} y^{(t)}
\end{equation} 
and \emph{perceived expertise} $\hat y^{(t)}(y^{(t)},\hat M_d^{(t)}) \in \Delta_n$ is defined as 
\begin{equation}
\label{eq:expertise-perceived}
    \hat y^{(t)}\big(y^{(t)},\hat M_d^{(t)}\big) = \big(\hat M_d^{(t)\top}y^{(t)}\big)^{-1}\diag\big(M_d^{(t)}\big)y^{(t)}.
\end{equation}
Then depending on the model, $x^{(t)}$ is equal to either $\bar y^{(t)}$ or $\hat y^{(t)}$. Our proposed models use also a scaling parameter $\tau\in(0,1)$ that can be adjusted to change the time-scale of the dynamics. If we have information on past reported influence matrices and expertise levels of team members, these models can be used to predict future influence matrices. Our three models are presented below.
\begin{itemize}
    \item \emph{Differentiation model (D model):} Motivated by Hypothesis~\ref{hp:hypothesis1}, this model assumes individuals assign influence based on the normalized, cumulative expertise~\eqrefreg{eq:expertise-normalized}, where individuals who perform better are accorded higher influence. The model is defined for all $i,j\in\{1,\dots,n\}$ as
    \begin{equation}
    \label{model:diff}
        \hat M_{ij}^{(t+1)} = (1-\tau) \hat M_{ij}^{(t)} + \tau \bar y_{j}^{(t)},
    \end{equation}
    which reads in matrix form as $\hat M^{(t+1)} = (1-\tau)\hat M^{(t)} + \tau \ones[n] \bar y^{(t)\top}$.

    \item \emph{Differentiation, Reversion model (DR model):} Motivated by Hypotheses~\ref{hp:hypothesis1} and \ref{hp:hypothesis2}, this model is based on the D model and assumes high-performing individuals are accorded more influence and low-performing individuals tend to assign influence weights uniformly amongst team members. The model uses the normalized expertise~\eqrefreg{eq:expertise-normalized} and is defined for all $i,j\in\{1,\dots,n\}$ as

    \begin{equation}
    \label{model:diff-rev}
    \hat M_{ij}^{(t+1)} = (1-\tau)\hat M_{ij}^{(t)} + \tau \Big( \bar y_{i}^{(t)} \bar y_{j}^{(t)} + \big(1-\bar y_{i}^{(t)}\big)\frac{1}{n}\Big),
    \end{equation}
    
    which reads in matrix form as 
    \begin{equation*}
        \hat M^{(t+1)} = (1-\tau)\hat M^{(t)} +  \tau\Big( \bar y^{(t)} \bar y^{(t)^\top} + \frac{1}{n}\big(\ones[n]-\bar y^{(t)}\big) \ones[n]^\top \Big).
    \end{equation*}
    
    \item \textbf{Cognitive model based on Differentiation, Reversion, Perceived expertise model (DRP model):} Motivated by Hypotheses~\ref{hp:hypothesis1}, \ref{hp:hypothesis2} and \ref{hp:hypothesis3}, this model is an extension of~\eqrefreg{model:diff-rev-skewed}, where everyone's expertise is misevaluated based on their own self-confidence. The model then uses the perceived expertise~\eqrefreg{eq:expertise-perceived} to learn how much influence is accorded to one another. Using all three hypotheses baked into this model provides the most accurate and consistent estimation.This model is defined for all
    $i,j\in\{1,\dots,n\}$ as
 
    \begin{equation}
    \label{model:diff-rev-skewed}
        \hat M_{ij}^{(t+1)} = (1-\tau)\hat M_{ij}^{(t)} + \tau \Big( \hat y_{i}^{(t)}\hat y_{j}^{(t)} + \big(1-\hat y_{i}^{(t)}\big)\frac{1}{n} \Big).
    \end{equation}

\end{itemize}

The following Lemma states that the D, DR, and DRP model are well-posed and the dynamics preserves row-stochasticity of the influence matrices. 
\begin{lemma}[Dynamic models preserve row-stochasticity]
\label{thm:indiv-invariantProps}
    Consider the D model~\eqref{model:diff}, DR model~\eqref{model:diff-rev}, and DRP model~\eqref{model:diff-rev-skewed} with $\tau\in(0,1)$ and $y^{(t)} = y = [0,1]^n$. If $\hat M^{(1)}$ is row-stochastic, then $\hat M^{(t)}$ remains row-stochastic for all $t\geq 1$ under the D and DR model. 
    
    If additionally, there exists at least one $i$ such that $\hat M_{ii}^{(1)} > 0$ and $y_i>0$, then the DRP model is well-posed for finite $t$ and $\hat M^{(t)}$ remains row-stochastic for all $t\geq 1$.
\end{lemma}

Given constant expertise $y$, it is clear that the affine models, D and DR, converge. In simulations (see \nameref{S2_Appendix}), we also observe that the DRP model exhibits convergence behaviors to a unique equilibrium. The next Lemma illustrates convergence of the DRP model for $y=c \ones[n]$ with $c>0$.
\begin{lemma}[Equilibrium and convergence of DRP model with uniform expertise]
\label{thm:indiv-eq}
    Consider the DRP model~\eqrefreg{model:diff-rev-skewed}. 
    Assume for $\tau\in(0,1)$, constant uniform expertise values $y^{(t)} = c\ones[n]$ with $c>0$, $\hat M^{(1)}$ row-stochastic, and that there exists at least one $i$ such that $\hat M_{ii}^{(1)} > 0$. Then $\lim_{t\to\infty}\hat M^{(t)} = \frac{1}{n}\ones[n]\ones[n]^\top$.
\end{lemma}
The proofs for these lemmas can be found in the \nameref{S1_Appendix}.

\subsubsection*{Proposed linear model}
Using machine learning models we can take advantage of all available data to estimate influence matrices. Combining text, connectivity network, expertise, and historical appraisals produce a multi-dimensional prediction model. We have $N$ teams of $n$ individuals that go through $T$ game rounds. To have a general format, we represent influence matrix for team $m$ at round $t$ by $\matrixSym{M}^{(m, t)}$. We represent all aforementioned features in matrix format (shown by $\matrixSym{X}^{(m, t)}_k$ for matrix feature $k$ of team $m$ at round $t$). We need to estimate the weight variables that maps the features to the influence matrix. In total, we assume there are $K$ matrix variables, shown as $\matrixSym{W}_k;\quad k=1 \text{ to } K$. Ergo, a convex objective function for estimating the influence matrix is defined as
\begin{align}\label{eq:generalized_convex_optimization_model}
\begin{split}
    \min\limits_{\substack{\matrixSym{W}_k:\; k = 1 \text{ to } K,\\
    B}} & \sum\limits_{m=1}^{N}\sum\limits_{t=1}^T \Big\lVert \sum\limits_{k=1}^{K} \matrixSym{X}^{(m, t)}_k \matrixSym{W}_k^T
    + \matrixSym{B} - \matrixSym{M}^{(m, t)}\Big\rVert_F^2\\
    & + \lambda \Big( \sum\limits_{k=1}^{K} \lVert\matrixSym{W}_k\rVert_{1, 1}
    + \lVert\matrixSym{B}\rVert_{1, 1}\Big),\\
    \text{Subject to}\;\; & \sum\limits_{k=1}^{K} \matrixSym{X}^{(m, t)}_k \matrixSym{W}_k^T + \matrixSym{B} \geq 0,\;\; \forall m \in [1, N],\;\forall t \in [1, T]\\
    & \ones[n]^T \Big(\sum\limits_{k=1}^{K} \matrixSym{X}^{(m, t)}_k \matrixSym{W}_k^T + \matrixSym{B}\Big) = \ones[n]^T.\;\; \forall m \in [1, N],\;\forall t \in [1, T]
\end{split}
\end{align}

\noindent where $\matrixSym{M}^{(m, t)}$ is the ground truth influence matrix, self-reported by team $m$ at round $t$. Depending on the application, for the history of the influence matrix, we may use only the first matrix, the average of all previous ones, or only the previous matrix. Variables to be calculated via optimization are the $n \times n$ weight matrices $\matrixSym{W}_k$. $\matrixSym{B}$ is the $n \times n$ bias matrix also to be estimated. We also use an $l1$-norm regularization to introduce sparsity to the estimated parameters that is commonly used in many real applications and also decrease the potential search space and therefore provides efficiency for the optimization solver.

\begin{lemma}
    The problem in~\eqref{eq:generalized_convex_optimization_model} is convex.
\end{lemma}

Based on the application, when only the probability distribution and the order of influence toward others is more important than exact values, we use cross-entropy as the loss function and KL divergence as the metric. In such a case, we can formulate the matrix estimation problem as the estimation of each row, which is a discrete distribution comprised of four numbers. Cross-entropy for two probability distribution of $p$ and $q$ is defined as $H(p, q) = -\sum\limits_{i=1}^{n}p_i\log q_i$. In this study, the two probabilities are 
\begin{align*}
    p &= \matrixSym{M}_{i,.} & \forall i \in [1, n]\\
    q &= \hat{\matrixSym{M}}_{i,.} = \sigma(\matrixSym{O}_{i,.}) = \sigma(\matrixSym{W}^T \matrixSym{X}_{i,.} + b) & \forall i \in [1, n]
\end{align*}
\noindent where $\sigma$ represents Softmax function, $M_{i,.}$ shows row $i$ of matrix $M$, and $W$ and $b$ show the weight and bias variables to be estimated.

\begin{lemma}
The optimization problem using the cross-entropy loss function on corresponding rows of two matrices $M$ and $\hat M$ can be written as
\begin{align}\label{eq:generalized_vectorized_convex_optimization_model}
    \begin{split}
        \min\limits_{\matrixSym{W}, b} &\quad -\sum\limits_{m=1}^{N}\sum\limits_{t=1}^{T} \sum\limits_{j=1}^{n} \sum\limits_{k=1}^{n} \matrixSym{M}_{j, k}^{(m, t)}\bigg( \matrixSym{X}_{j, k}^{(m, t)} \matrixSym{W}_{k, j} + b_j \\
        &  - \log\sum\limits_{l=1}^{n} \exp\big( \matrixSym{X}_{j, k}^{(m, t)} \matrixSym{W}_{k, j} + b_j \big) \bigg) + \lambda\bigg(\|\matrixSym{W}\|_1 + \|b\|_1\bigg)
    \end{split}
\end{align}
\end{lemma}

The problem in~\eqref{eq:generalized_vectorized_convex_optimization_model} does not require any constraints to solve the convex optimization. First, because the Softmax function ($\sigma$) in this equation provides a discrete distribution in the format of vectors (all fall into $[0, 1]$ and sum up to 1). Second, since here we format the data points as vectors and not as matrices.

\begin{lemma}
    The problem in~\eqref{eq:generalized_vectorized_convex_optimization_model} is convex.
\end{lemma}

The proofs for all these lemmas can be found in the \nameref{S1_Appendix}.

\subsubsection*{Proposed deep neural network-based model}
\begin{figure}[!ht]
    \centering
    \includegraphics[width=\linewidth]{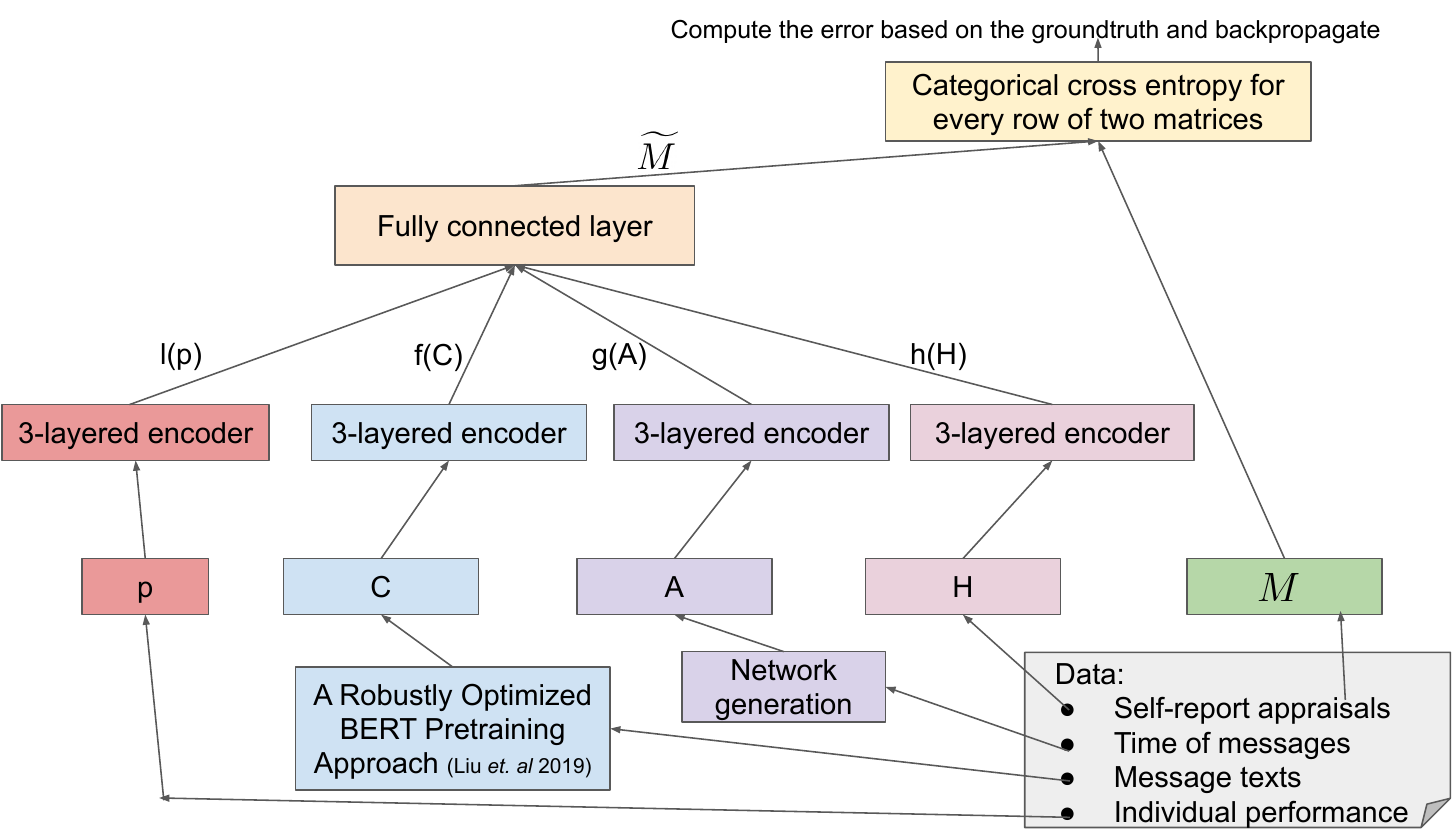}
    \caption{\textbf{Deep learning model architecture:} A deep encoder model in a two-tower framework~\cite{he2017neural} for learning the three mappings of connectivity network, content of messages, and history of appraisals. The final layer computes the cosine similarity with the ground truth influence matrix and back-propagates the error using Stochastic Gradient Descent (SGD).}
    \label{fig_deep-learning-model-architecture}
\end{figure}

We can also learn the mapping defined by the three weight matrices as deep encoders in a two-tower model~\cite{he2017neural}. In this regard, we apply end-to-end models to estimate the social influence matrices using multi-layered encoders from raw features to an influence matrix. This is described in Fig.~\ref{fig_deep-learning-model-architecture}. Each encoder is comprised of three fully connected Exponential Linear Unit (ELU)~\cite{clevert2015fast} layers, initialized by He \emph{et al.}~\cite{he2015delving} Normal initialization, such that it draws samples from a truncated normal distribution centered on 0 with a standard deviation of $\sqrt{\frac{2}{f}}$ where $f$ is the number of input units in the weight tensor. We use Dropout~\cite{srivastava2014dropout} after each fully connected layer to decrease overfitting. Then, all three outputs are concatenated and fed to another three fully connected layers with the same activation function to decrease the dimensionality of embedding vectors to an $n \times n$ matrix $\tilde{M}^{(m, t)}$, $n$ being the number of individuals. Finally, cosine similarity of the two matrices $M^{(m, t)}$ and $\tilde{M}^{(m, t)}$ is computed and the error is back-propagated using stochastic gradient descent.

The deep method description is shown in Fig.~\ref{fig_deep-learning-model-architecture}. In this figure, the weights matrices in~\eqref{eq:generalized_convex_optimization_model} are framed as a layers of deep neural networks. This model creates a non-convex problem; however, arguably with the abundance of data, a more effective model.

\subsection*{Baseline dynamical models}
Below, we provide an overview of our baseline dynamical models used to estimate the influence matrix evolution.
\begin{itemize}
    \item \emph{Constant appraisal model:} This is a basic model which assumes the influence matrix remains constant over time. The model reads in matrix form as $\hat M^{(t+1)} = M^{(t)}$, or defined element-wise, for all $i,j\in\{1,\dots,n\}$ as
    \begin{equation}
    \label{model:baseline}
        \hat M_{ij}^{(t+1)} = \hat M_{ij}^{(t)}.
    \end{equation}
    
    \item \emph{Reflected appraisal model:} The reflected appraisal model is based on the model proposed in Mei \emph{et al.}~\cite{mei2017dynamic}. The self-influence estimate $\hat{M}^{(t+1)}_{ii}$ increases relative to $\hat M_{ij}^{(t)}$ if the expertise of individual $i$, $y_i^{(t)}$, is larger then average team expertise observed by individual $i$, $\sum_{i=1}^{n}\hat M_{ik}^{(t)}y_k^{(t)}$. If $\hat{M}^{(t)}_{ii}$ increases, then the interpersonal weights $\hat{M}^{(t+1)}_{ij}$ for all $j\neq i$ are decreased so that $\hat{M}^{(t+1)}$ remains row-stochastic. 
    The reflected appraisal model is defined element-wise for all $i,j\in\{1,\dots,n\}$ and $t>0$ as
    \begin{equation}
    \label{model:reflected-appraisal-element}
        \begin{split}
            \hat{M}_{ii}^{(t+1)} &= \hat M_{ii}^{(t)}
             + \hat M_{ii}^{(t)} (1 - \hat M_{ii}^{(t)}) \Big( \bar y_i^{(t)} - \sum_{k=1}^{n} M_{ik}^{(t)}\bar y_k^{(t)} \Big), \\
            \hat{M}_{ij}^{(t+1)} &= \hat M_{ij}^{(t)} 
            - \hat M_{ii}^{(t)} \hat M_{ij}^{(t)} \Big( \bar y_i^{(t)} - \sum\limits_{k=1}^{n}\hat M_{ik}^{(t)}\bar y_k^{(t)} \Big),
        \end{split}
    \end{equation}
    which reads in matrix form as
    \begin{equation*}
    \label{model:reflected-appraisal-matrix}
        \hat{M}^{(t+1)}  = \hat M^{(t)} +
        \diag\big( (I_n-\hat M^{(t)})\bar y^{(t)}\big) \diag(\hat M_d^{(t)}) (I_n - \hat M^{(t)}).
    \end{equation*}
    
    \item \emph{Structural balance theory:} Structural balance theory is a long-established theory describing the dynamics that govern the sentiment of interpersonal relationships. Researchers have consistently delivered various theoretical~\cite{marvel2011, srinivasan2011local, facchetti2011computing, heider1946attitudes} and empirical support~\cite{friedkin2019positive, nc2019, szell2010} for the emergence of this phenomenon in myriad settings. In this study, we use a generalized Structural balance theory model (SBT) that is inspired by earlier research Kulakowski \emph{et al.} ~\cite{kulakowski2005heider}. It predicts the dynamic of influence as introduced in the following
    \begin{align}
        \hat{M}^*_{ij} &= \frac{1}{n-2} \sum_{k=1,\; \text{when } k \neq i,\; k \neq j}^{n} \hat M^{(t)}_{ik} \hat M^{(t)}_{kj},\\
        \hat{M}^{(t+1)}_{ij} &= \hat{M}^*_{ij} / \ones^\top \hat{M}^*_{i,.}
    \end{align}
\end{itemize}


\section*{Results}

\subsection*{Origins of interpersonal influence}
The following hypotheses, motivated from past research, are empirically supported by our experimental results:

\begin{hyp} \label{hp:hypothesis1}
Individuals with higher expertise are accorded higher interpersonal influence from the group.
\end{hyp}

\begin{hyp} \label{hp:hypothesis2}
Individuals with lower expertise have diminished ability to recognize experts in the group.
\end{hyp}

\begin{hyp} \label{hp:hypothesis3}
Individuals with higher confidence are accorded higher interpersonal influence from the group.
\end{hyp}

In this experiment, subjects read and answer every question individually. Ergo, the individual performance (expertise) can be measured by the ratio of correct answers one gives individually, prior to seeing others' answers and the discussion phase. Assuming individuals can potentially keep track of others' expertise by recalling their answers or their chat messages, we study if individual expertise plays a prominent role in the amount of social influence one receives. Note that $\matrixSym{M}$ shows the ground truth influence matrix.

\begin{figure}[!ht]
    \centering
    \includegraphics[width=\linewidth]{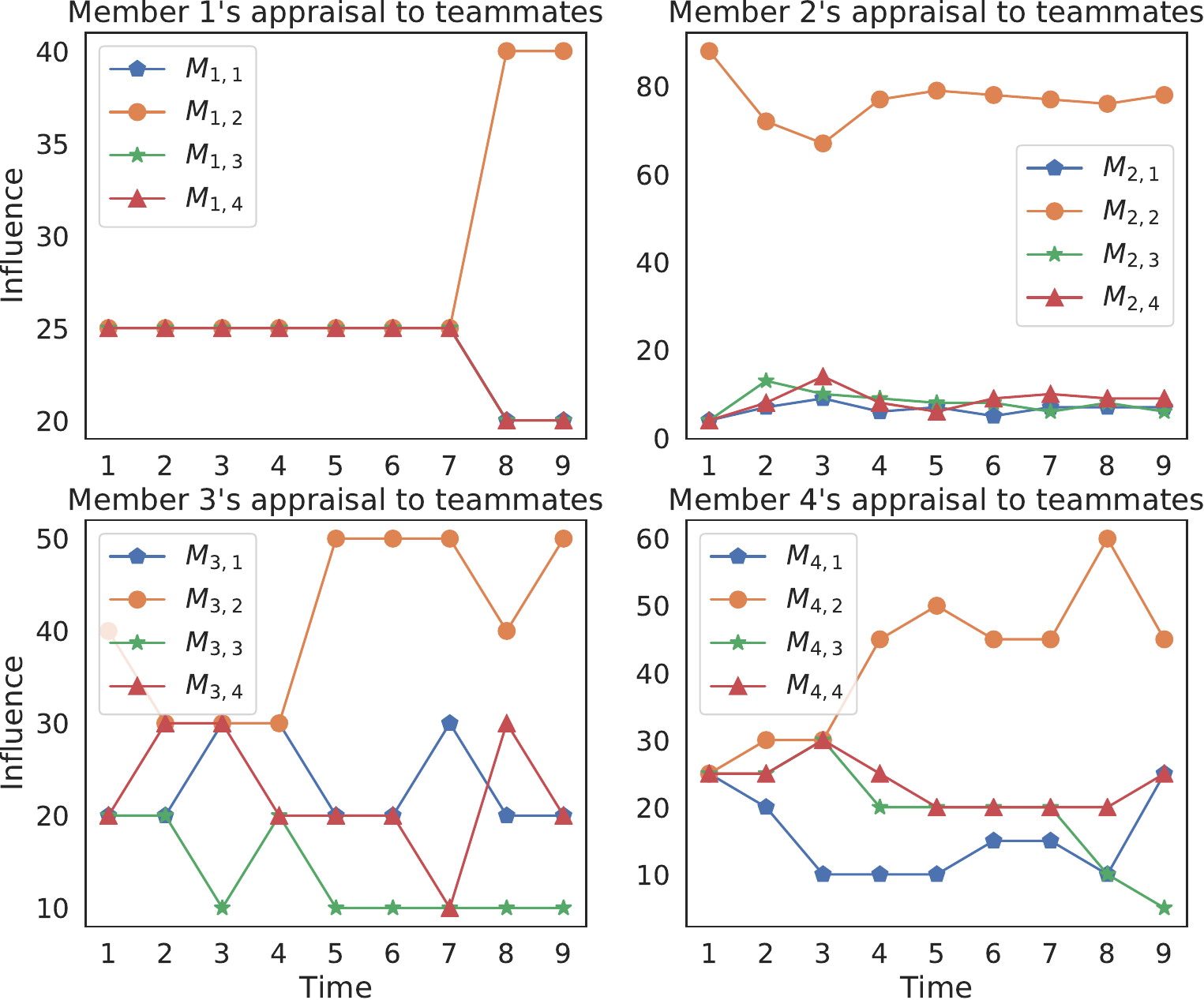}
    \caption{\textbf{Dynamics of the influence matrix in one team:} $M$ shows a $4\times4$ influence matrix for this team. Every panel shows how much every subject reports others influenced them over time. in other words, it shows the amount of appraisal every person assigns to team members including themselves over time. After answering all questions, we observe that member 2 is the most accurate (cumulative correctness rate for member \#1= 49\%, member \#2= 70\%, member \#3= 36\%, and member \#4= 58\%). This figure illustrates the team's interpersonal appraisals reflect the accuracy of the team members, which was ascertained early on in the experiment.}
    \label{fig_dynamics-of-the-influence-matrix-in-one-team}
\end{figure}

Qualitatively, we can look at the dynamics of every subject's appraisal of every subject over time. For example, Fig.~\ref{fig_dynamics-of-the-influence-matrix-in-one-team} show a team of four subjects and $\matrixSym{M}^{(t)}_{ij}$ shows the amount of influence subject $i$ assigns to subject $j$ in every round. It is clear that on aggregate and over time the individuals found member \#2 to be the most accurate and therefore they reported member \#2 to be the most influential person. Also, we know after answering all questions, member \#2 was more accurate than anybody else (cumulative correctness rate for member \#1= 49\%, member \#2= 70\%, member \#3= 36\%, and member \#4= 58\%). This is an example of the emergence of the first hypothesis. This figure illustrates how the interpersonal appraisal reflects the underlying expertise and how teammates were able to uncover that expertise early on in the experiment.

To quantitatively test the three hypotheses, we define a few terms based on the influence matrix: confidence, persuasiveness, and mean reversion. The definitions of social confidence and persuasiveness for person $i$ take into account column $i$ of the influence matrix. Confidence is provided in~\eqref{eq:local_confidence}. The definition for persuasiveness is~\eqref{eq:local_persuasiveness}. To compute persuasiveness, we consider the reflective relative appraisal matrix $\matrixSym{C}$, which is defined by removing the diagonal elements from matrix $\matrixSym{M}$ and re-scaling to be row-stochastic. Another term defined in the following is expertise that is the proportion of questions every member individually answers correctly.

\begin{itemize}
    \item \emph{Expertise}: Individual correct answer rate (individual accuracy or individual performance) --- the proportion of questions one has individually answered correctly up to any given time ($\frac{\text{\#correct answers}}{\text{\#answers}}$).

    \item \emph{Confidence}: Self-appraisal --- the amount of influence one assigns to oneself at any given time.

    \begin{equation} \label{eq:local_confidence}
        \matrixSym{M}^{(t)}_{ii}.
    \end{equation}

    \item \emph{Persuasiveness}: Appraisal by others --- the amount of influence everybody else is assigning to a particular team member at any given time. 
    \begin{equation}            \label{eq:local_persuasiveness}
        \frac{1}{n-1}\sum\limits_{j, \; j \neq i} \matrixSym{M}^{(t)}_{ji}.
    \end{equation}
    
    
    \item \emph{Mean reversion}: Reversion to the mean (uniformity) in the appraisals that an individual holds of their teammates:
    \begin{equation}\label{eq:reversion}
      D_i = \sum_{j=1}^{n} |M_{ij} - \frac{1}{n}|^2.
    \end{equation}
\end{itemize}

The empirical distribution of the expertise shows that individuals are more accurate than a random guess (on average individuals have about 50\% correct rate when the expected value is 25\% for four-choice questions). Also, we find controlling for the difficulty of the questions does not change the cumulative correctness rate distribution.


\subsubsection*{Regression study of interpersonal influence}
In this section, we intend to estimate one's average influence reported by their teammates at the end of the experiment using Regression models. We use the Generalized Linear Model method to solve the regression problem. Table~\ref{tbl:regression_coefficients} shows the coefficients and their statistical significance in three least-squares problems (each column shows a separate test).

\begin{table*}[!ht]
\begin{adjustwidth}{-1.75in}{0in} 
\centering
\begin{tabular}{|l|l|l|l|l|}
\hline
\textit{Predicting Persuasiveness} & Feature-set 1 & Feature-set 2 & Feature-set 3 & Feature-set 4 \\ \hline
Intercept & 0.13 *** & 0.19 *** & 0.12 *** & 0.12 *** \\ \hline
Expertise & \textbf{0.20 ***} &  & \textbf{0.17 *** (VIF: 1.05)} & \textbf{0.17 *** (VIF: 1.11)} \\ \hline
Confidence &  & \textbf{0.14 ***} & \textbf{0.11 *** (VIF: 1.05)} & \textbf{0.12 *** (VIF: 1.06)} \\ \hline
Response network out-degree &  &  &  & 0.00 (VIF: 2.63) \\ \hline
Sentiment network out-degree &  &  &  & 0.002 ** (VIF: 2.54) \\ \hline
 & \begin{tabular}[c]{@{}l@{}}Log-likelihood:  162.86\\ AIC:  -321.7\\ BIC:  -316.3\end{tabular} & \begin{tabular}[c]{@{}l@{}}Log-likelihood:  162.90\\ AIC:  -321.8\\ BIC:  -316.4\end{tabular} & \begin{tabular}[c]{@{}l@{}}Log-likelihood:  168.12\\ AIC:  -330.2\\ BIC:  -322.1\end{tabular} & \begin{tabular}[c]{@{}l@{}}Log-likelihood:  170.18\\ AIC:  -330.4\\ BIC:  -316.8\end{tabular} \\ \hline
\end{tabular}
\caption{\textbf{Regression result for predicting persuasiveness:} Generalized Linear Model regression coefficients and their statistical significance in estimating local persuasiveness at the end of the experiment (after answering 45 questions). The statistical significance using the $p$-value is portrayed with: *** for $p<0.01$ and ** for $p<0.05$. The amount of Variance Inflation Factor (VIF) is provided in parenthesis; this factor estimates how much the variance of a regression coefficient is inflated due to multicollinearity in the model. Statistical results remain significant in models with multicorrelated independent variables when VIF $< 5$~\cite{everitt2002cambridge}. Taking into account the interactions of all variables, we find that expertise and confidence are consistently statistically predictive of persuasiveness. Networks are extracted from the time and the content of chat messages among individuals and defined in the "Feature-set from logs" part.}
\label{tbl:regression_coefficients}
\end{adjustwidth}
\end{table*}
The empirical evidence for Hypothesis~\ref{hp:hypothesis2} is obtained via regression on expertise and mean reversion. Table~\ref{tbl:reversion_regression} shows the regression results for predicting mean reversion for every individual. Our results show the more expert one individual is, the more diversely they appraise their teammates, as expertise is predictive of the reversion to the mean (with positive coefficient and $p$-value $< 0.05$) in all teams. It also shows this is an individual feature as the team average performance is not statistically significant while individual performance stays statistically significant. In fact, this result is the motivation behind using expertise as a weighted average in the cognitive dynamical model described in the next section.

 \begin{table}[!ht]
\centering
\begin{tabular}{|l|l|l|}
\hline
\textit{Predicting Mean reversion} & Feature-set 1 & Feature-set 2 \\ \hline
Intercept & 0.10 *** & 0.10 *** \\ \hline
Individual performance & \textbf{0.07 **} & \textbf{0.08 ** (VIF: 1.21)} \\ \hline
Team performance &  & -0.004 (VIF: 1.21) \\ \hline
 & \begin{tabular}[c]{@{}l@{}}Log-likelihood:  384.4\\ AIC:  -764.8\\ BIC:  -754.4\end{tabular} & \begin{tabular}[c]{@{}l@{}}Log-likelihood:  384.4\\ AIC:  -762.8\\ BIC:  -747.2\end{tabular} \\ \hline
\end{tabular}
\caption{\textbf{Regression result for predicting mean reversion:} The statistical significance using the $p$-value is portrayed with: *** for $p<0.01$ and ** $p<0.05$. The amount of Variance Inflation Factor (VIF) is provided in parenthesis; this factor estimates how much the variance of a regression coefficient is inflated due to multicollinearity in the model. Statistical results remain significant in models with multicorrelated independent variables when VIF $< 5$~\cite{everitt2002cambridge}. This result shows the individual performance (expertise) is positively correlated with the reversion to the mean.}
\label{tbl:reversion_regression}
\end{table}
Table~\ref{tbl:regression_coefficients} shows that introducing more variables in columns has increased log-likelihood, Bayesian Information Criterion (BIC), and Akaike Information Criterion (AIC). It shows that expertise has a consistently positive and statistically significant impact on persuasiveness (the empirical evidence for Hypothesis~\ref{hp:hypothesis1}). This statistical significance is robust even after adding many more variables shown in the rightmost column. Also, confidence has a positive and statistical effect to predict persuasiveness (the empirical evidence for Hypothesis~\ref{hp:hypothesis3}). However, its coefficient (importance) is less than expertise (aligned with research in confidence heuristics~\cite{thomas1995confidence}). This result is found when the platform provides immediate feedback for every question. If no feedback is provided or there is no right or wrong answer (i.e. judgmental questions), people might use confidence as a more substantial metric in their appraisal distribution.


\subsubsection*{Causality study of interpersonal influence}
To study the order of the effects by confidence, persuasiveness, and expertise and to what extent their effect is supported by data, we propose to use a forecasting causality test. We use Granger causality --- a statistical concept of causality that is based on prediction~\cite{granger1969investigating, granger1980testing, granger2001essays}. To study the causality of the aforementioned variables, we compute expertise, confidence, and persuasiveness in every round. Thus, in this data, for every person, we have three time series with nine data points. For every person, we compute Granger causality of these time series and study what percentage of change in individuals' expertise, confidence, and persuasiveness have statistical causal effects. Note that the exact number of lags, which were statistically significant, could uncover the precedence of these variables. That plausibly opens the door to studying the order of different effects, such as hypotheses 1 and 2, leading to social influence.

\begin{figure}[!ht]
    \centering    \includegraphics[width=0.5\linewidth]{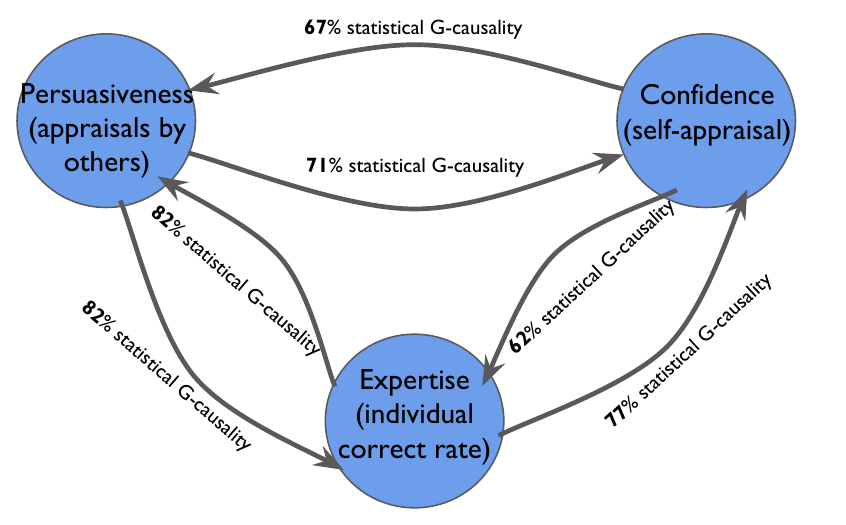}
    \caption{\textbf{Granger causality result:} This figure shows the proportion of statistically significant Granger causality of timeseries of confidence, persuasiveness and expertise in all teams. The $p$-values have been corrected using Benjamini-Hochberg (BH) procedure with False Discovery Rate of 5\% has required $p\text{-value} < 0.03$ as statistical significance threshold.}
    \label{fig_granger-causality-result}
\end{figure}

Applying statistical tests on three time series per individual is susceptible to false discovery. Ergo, in the following results, obtained $p$-values are adjusted using Benjamini-Hochberg (BH)~\cite{benjamini1995controlling} controlling procedure with false discovery rate (FDR) of 5\%. Fig.~\ref{fig_granger-causality-result} depicts empirical evidence for Hypothesis~\ref{hp:hypothesis1} and \ref{hp:hypothesis3}. It shows multiple Granger causality tests on the effect of confidence, persuasiveness, and expertise on one another. These results are obtained after applying the BH procedure with FDR=5\% that has $p\text{-value} < 0.03$ as the statistical significance threshold. This result shows that in most individuals there is statistical causation from confidence to persuasiveness and vice versa over time. The results also show there is a causal relationship from expertise to confidence (aligned with confidence heuristics) and also persuasiveness (TMS~\cite{lewis2003measuring}). Note that unlike the regression experiment that takes into account only the final influence report and expertise, causality results are computed from the time series of influence and performance for every individual during the course of the experiment.

\subsection*{Influence matrix estimation}
Studying the problem of estimating the influence matrix is novel and challenging. Our data present an unprecedented opportunity to understand team behavior and estimate the interpersonal influence system of teammates. Note that $\matrixSym{M}$ represents the ground truth influence matrix, and $\hat{\matrixSym{M}}$ represents the estimated influence matrix.

The Mean Square Error (MSE) and Kullback-Liebler (KL) divergence of two row-stochastic matrices $\matrixSym{M}$ and $\hat{\matrixSym{M}}$ with $n$ rows are defined as follows,
\vsb
\begin{equation}\label{eq:mse_error_def}
    \text{MSE}(\matrixSym{M}, \hat{\matrixSym{M}}) = \tfrac{1}{n} \| \matrixSym{M} - \hat{\matrixSym{M}}\|_F^2 = \tfrac{1}{n} \sum_{i=1}^{n}\sum_{j=1}^{n} |M_{ij} - \hat M_{ij}|^2,
\end{equation}
\vsb
\begin{equation}\label{eq:kld_def}
    \text{KL}(\matrixSym{M}, \hat{\matrixSym{M}}) 
    = \tfrac{1}{n}\! \sum_{i=1}^n \!\bigg(\! \sum_{j=1}^n M_{ij} \log\! \frac{M_{ij}}{\hat M_{ij}} \bigg).
\end{equation}

Depending on the application, one may choose any of the above metrics. To showcase the generality of our proposed models, we present results on both metrics.

We propose three models with a spectrum of explainability: (i) A black-box deep learning model which is the most accurate, (ii) A white-box linear model using convex optimization that explains substantial features leading to interpersonal influence
, (iii) A cognitive dynamical model which postulates an underlying mechanism. This dynamical model unlike the other two machine learning models does not require much training data as it only has one scalar hyperparameter to choose from. In the following, we introduce the results of estimation using these models.

\subsubsection*{Influence matrix estimation: cognitive dynamical models}
We propose discrete-time dynamical models, of the form $\hat{M}^{(t+1)} = T(\hat{M}^{(t)},x^{(t)})$, that are formulated such that established sociological concepts are baked into their equations. These non-machine learning models only take the history of past local influence weight information and individuals' performance values. These models can be used to provide a single or multi-round forecast of the influence matrices for successive rounds of the experiment. Our results compare the accuracy of various dynamical models, which are described above in the Materials and Methods section. The models are a baseline model, a model based on Hypothesis~\ref{hp:hypothesis1}, a model based on Hypotheses~\ref{hp:hypothesis1} and \ref{hp:hypothesis2}, and a model based on Hypotheses~\ref{hp:hypothesis1}, \ref{hp:hypothesis2} and \ref{hp:hypothesis3}. The models assume that individuals can observe each other's expertise, which they take into account when readjusting the influence weights assigned to one another.

\begin{figure*}[!ht]
    \centering
    \includegraphics[width=\linewidth]{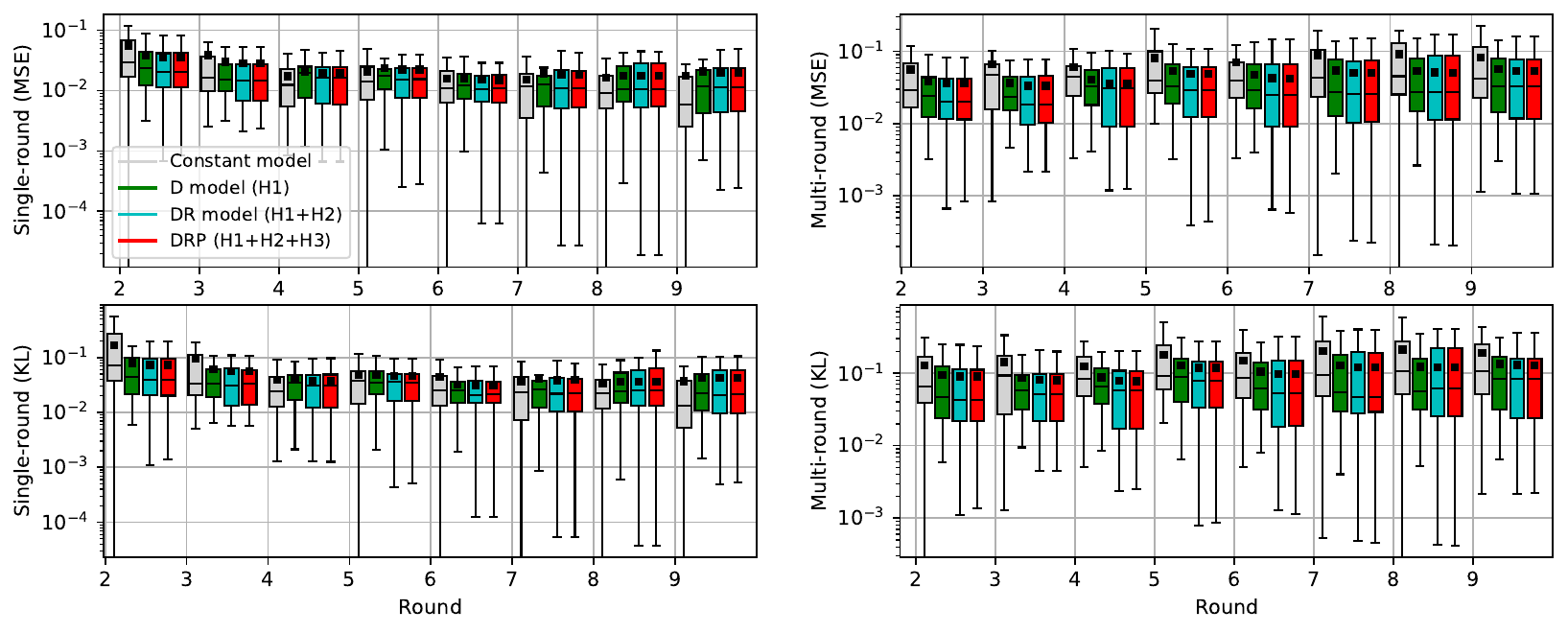}
    \caption{\textbf{Cognitive model evaluation:} The mean squared error (MSE) and the Kullback-Leibler (KL) divergence for different dynamical models over nine rounds of influence matrix estimation. Differentiation (D model) takes into account hypothesis~\ref{hp:hypothesis1}. Differentiation, Reversion (DR model) is inspired by hypotheses~\ref{hp:hypothesis1} and~\ref{hp:hypothesis2}. Differentiation, Reversion, Perceived (DRP model) uses hypotheses~\ref{hp:hypothesis1},~\ref{hp:hypothesis2}, and~\ref{hp:hypothesis3}. For the models, we use the hyperparameter $\tau=0.4$. In this figure boxes show the interquartile range of the errors, the whiskers show minimum and maximum of the range of the distribution. In each box, the dot shows the average and the line shows the median of the portrayed distribution.
    \textbf{Left:} Single-round forecast error of various dynamical models for predicting the influence matrix one round ahead. The models estimate $\hat{M}^{(t+1)}$ using the expertise $\bar y^{(t)}$ and the reported influence matrix from the previous round $M^{(t)}$.
    \textbf{Right:} Multi-round forecast error of various dynamical models for predicting the influence matrix multiple rounds ahead. Outliers are not shown for better readability. The models estimate $\hat{M}^{(t+1)}$ using the expertise $\bar y^{(t)}$ and the initial ground truth $M^{(1)}$ influence matrix reported by individuals. For rounds $t+1\geq 2$, the dynamics use the predicted influence matrix from the previous round $\hat{M}^{(t)}$, instead of $M^{(t)}$.
    For rounds $t\geq 4$, the influence network remains relatively constant, so the cognitive dynamical model offers incremental improvements to the baseline models for single-round forecast. However, this model gives significant improvements in accuracy from baseline models for all rounds in multi-round prediction.}
\label{fig_cognitive-model-evaluation}
\end{figure*}

We consider single-round and multi-round forecast, to compare how the models perform when using the previous round's influence matrix versus only the initial round's influence matrix. The \emph{single-round forecast} predicts the influence matrix at future rounds $t+1\geq2$ using the reported influence matrix from the previous round $M^{(t)}$ and the expertise $y^{(t)}$. Note that for the single-round forecast, the prediction comes from the following modification to the dynamics, $\hat M^{(t+1)}=T(M^{(t)},y^{(t)})$. 
Fig.~\ref{fig_cognitive-model-evaluation} (left) illustrates the error of a given model. The \emph{multi-round forecast} predicts $\hat M^{(t)}$, for any future round $t+1\geq 2$, using the initial reported influence matrix $M^{(1)}$ and the previous round's expertise values $y^{(t)}$ as inputs. The following details how the dynamical models are modified to give multi-round forecasts. To estimate $\hat M^{(2)}$, the map $T(M^{(1)},y^{(1)})$ is used. For subsequent rounds $t\geq 3$, the estimate comes from $\hat M^{(t+1)}=T(M^{(t)},y^{(t)})$. In summary, the ground truth influence matrix data is propagated over a sequence of rounds to predict the influence matrix at future rounds. Fig.~\ref{fig_cognitive-model-evaluation} (right) illustrates the error of a given model.

Overall for the single and multi-round forecast, we observe increased estimation accuracy for the models that capture more hypotheses. For the single-round forecast, we observe that the accuracy increases for later rounds since individuals adjust influence weights less as the experiment goes on. However, the accuracy for later rounds does not give significant improvements compared to the constant baseline model, since the influence weights remain relatively constant for rounds $t\geq4$. For the multi-round forecast, as expected, we see that the accuracy decreases for predictions of later rounds; yet consistently provides the most accurate predictions of the influence matrices regardless of whether the model is given the most up-to-date ground truth values. In the following, we also show that the cognitive dynamical model gives competitive predictions compared to the machine learning models.

\subsubsection*{Influence matrix estimation: machine learning models}
We introduce machine learning models to predict the influence matrix at every round. These powerful models are able to take multiple features extracted from the logs of the experiment and learn a mapping to estimate the corresponding influence matrix. In order to learn such mappings, they require training data. Thus, we use a portion of collected logs as the training and apply the trained model on the unseen logs. 

In order to predict the influence matrix at round $t$, we use time and content of text messages from the broadcast communication logs until round $t$, individual correct percent until round $t$, and reported influence matrices before round $t$ in the following.

We propose a linear maximum likelihood estimation model using convex optimization and a deep neural network model. Fig.~\ref{fig_deep-learning-model-architecture} shows the architecture of the neural network model (see Methods and Materials). They both similarly intend to find a linear or nonlinear combination of the aforementioned input features to estimate a row-stochastic influence matrix. We compare the proposed models to the baseline models with a variation of different input features. All models are trained with 80\% of the data and tested on the withheld 20\%. To compute the statistical significance, we draw 1000 bootstraps with replacement from the hold-out test set.

\begin{figure*}[!ht]
    \centering
    \includegraphics[width=\linewidth]{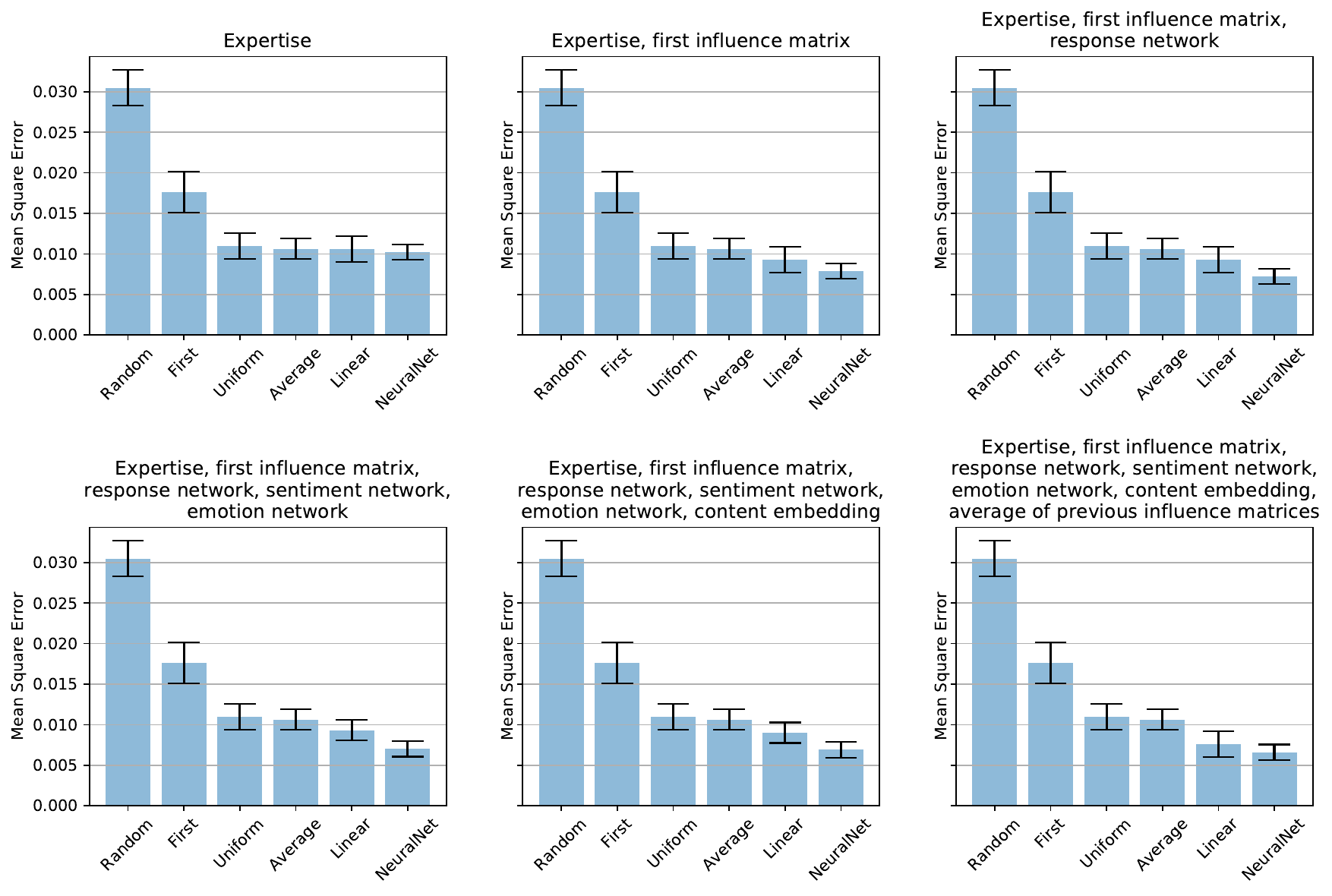}
    \caption{\textbf{Improvement in machine learning models by adding more features:} Average of Mean Square Error (MSE) of estimated influence matrix from the ground truth in the test set of influence matrices. Titles show the list of features fed to the models. The error bar shows the standard deviation in 1000 bootstrap on test error. Both machine learning models improve when given more features from the logs.}
    \label{fig_mse-models-with-features}
\end{figure*}

Due to the application, for every round, we assume only the first influence matrix is given, and we need to predict all influence matrices in future rounds with that. Hence, in Fig.~\ref{fig_mse-models-with-features}, we use only the first influence matrix with expertise, text embeddings, and so forth, in every round, to predict a $4 \times 4$ influence matrix. It shows the average of Mean Square Error (MSE) for the estimated influence matrix from the ground truth influence matrix (reported by individuals) in every round for every team. MSE is defined by~\eqref{eq:mse_error_def}. Fig.~\ref{fig_mse-models-with-features} depicts linear and neural network-based models consistently surpass other baselines with any sets of features. Also, it shows the proposed models become more accurate as more features are introduced.

\begin{figure*}[!ht]
    \centering
    \includegraphics[width=\linewidth]{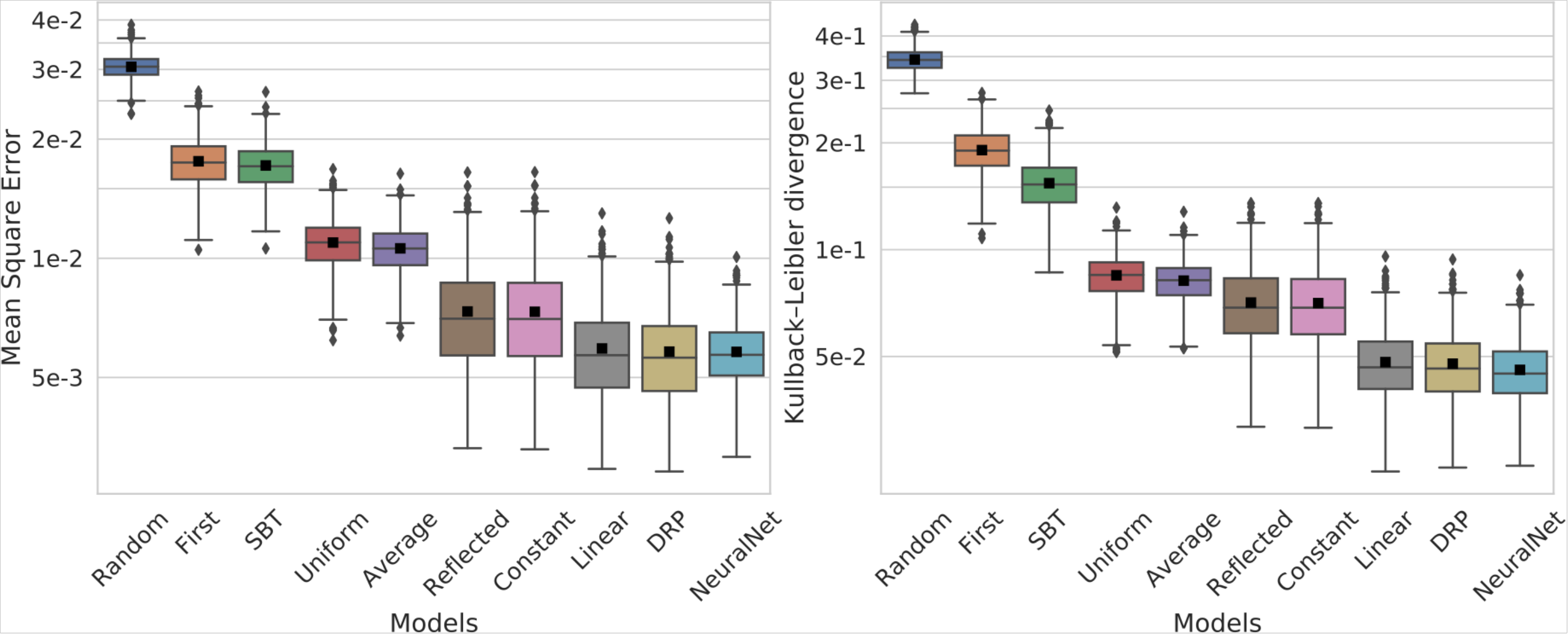}
    \caption{\textbf{Comparison of all models:} Mean squared error (MSE) and Kullback–Leibler (KL) divergence of single-round influence matrix prediction for baseline algorithms and the proposed models. Evaluations are applied on 1000 bootstraps of the holdout test dataset (20\% of the entire data). All models have access to the expertise and previous influence matrix for every team. The box shows the interquartile range of the errors, the whisker shows minimum and maximum of the range of the distribution, and the dots show the outliers. Baseline models: {\em Random} baseline is a randomly generated row-stochastic matrix. {\em First} predicts a team's first influence matrix to be unchanged. {\em SBT} baseline uses the generalized Structural Balance Theory~\cite{kulakowski2005heider}. {\em Uniform} predicts a matrix with all elements as $\frac 1 n$. {\em Average} predicts a team's row-stochastic average influence matrix to be the most accurate prediction for any influence matrix. {\em Reflected} baseline uses reflected appraisal mechanism for prediction~\cite{mei2017dynamic}. {\em Constant} predicts the influence matrix to be unchanged from last measured one. Proposed models: Cognitive model based on Differentiation, Reversion, Perceived expertise model {\em DRP} takes into account the aforementioned hypotheses (Hypotheses~\ref{hp:hypothesis1}, \ref{hp:hypothesis2}, \ref{hp:hypothesis3}) to predict influence matrices. Moreover, Linear model using convex optimization ({\em Linear}) and Neural Networks model ({\em NeuralNet}) are proposed to learn important features from the logs to estimate influence matrices. The figure depicts the proposed models outperform baselines. Surprisingly, this figure shows the reflected appraisal model does not surpass the baseline of considering previous influence matrix to be unchanged (Constant model). This figure also shows that the cognitive model works competitively with the learning models showing the power behind our empirically proven hypotheses.}
\label{fig_comparison-of-all-models}
\end{figure*}

In another setting, for every round, we assume we have access to the previous influence matrix and expertise to predict the current influence matrix. For this setting, there are more baselines that we can compare our proposed models against. Fig.~\ref{fig_comparison-of-all-models} (left) shows MSE divergence error for models using previous influence matrix and expertise. Similarly, the neural network-based model surpasses all baselines and provides statistically significant lower MSE. It is worth mentioning that proposed linear model (\eqref{eq:generalized_convex_optimization_model}) is competitive with the proposed neural network model (Fig.~\ref{fig_deep-learning-model-architecture}). Also, interestingly, the proposed cognitive dynamical model (\eqref{model:diff-rev-skewed}) which does not require any training and is described by a mechanism that postulates past research in social psychology works significantly better than other baselines and competitively close to the proposed machine learning models. Note that both machine learning models use optimization methods for training that requires several steps to converge.

Fig.~\ref{fig_comparison-of-all-models} shows MSE and KL divergence of the estimated influence matrix from the ground truth reported by individuals. MSE is defined on two matrices in~\eqref{eq:mse_error_def} and KL divergence in~\eqref{eq:kld_def}. MSE and KL divergence provide two different perspectives regarding the efficacy of influence estimation. MSE formulation emphasizes the exact values in matrices, while KL divergence attends to the discrete probability distribution in corresponding rows of two matrices. Fig.~\ref{fig_comparison-of-all-models} shows that for both error measurements, the two proposed models work efficiently compared to all baselines with the neural network-based model surpassing all other models. The problem formulation for MSE of influence matrices is provided in~\eqref{eq:generalized_convex_optimization_model}, and for probability distribution estimation in every row of influence matrices is given in~\eqref{eq:generalized_vectorized_convex_optimization_model} (see Methods and Materials).

Table~\ref{tbl:features_importance} sheds light on the importance of every feature. This table shows entry-wise l1-norm of estimated parameters in~\eqref{eq:generalized_convex_optimization_model}. The values in table~\ref{tbl:features_importance} are sorted from most to least important top to bottom. This result shows the previous influence matrix is the most important feature used to predict the next influence matrix, while expertise is the second most predictive, and text embedding the third. It shows sentiment, emotion, and responsiveness networks are far less related to the social influence.


\begin{table}[!ht]
\centering
\begin{tabular}{|l|c|}
\hline & $l1$-norm of estimated parameters \\ \hline
Previous influence matrix & 0.2137 $\pm$ 0.0027 \\ \hline
Expertise & 0.0239 $\pm$ 0.0027 \\ \hline
Message content embedding & 0.0111 $\pm$ 0.0003 \\ \hline
Message sentiment & 0.0078 $\pm$ 0.0010 \\ \hline
Message emotion & 0.0050 $\pm$  0.0007 \\ \hline
Message responsiveness & 0.0041 $\pm$ 0.0004 \\ \hline
\end{tabular}
\caption{\textbf{Importance features in predicting influence:} Entry-wise $l1$-norm of estimated parameter matrix in linear model given in~\eqref{eq:generalized_convex_optimization_model}. This table shows the importance of each features in the proposed linear model. The embeddings~\cite{liu2019roberta}, sentiments, and emotions all are computed from the message text content; however, responsiveness is computed from the timestamps of the messages.}
\label{tbl:features_importance}
\end{table}

\section*{Discussion}
Interpersonal appraisal networks can be modeled as an influence matrix, where weighted edges signify positive or negative appraisals among people. Being able to estimate these influence matrices has important applications such as marketing advertisements, creating successful political campaigns, and improving the efficiency of communication among team members. 
The problem of influence matrix estimation within groups is novel; it has been studied previously either with simulated data or with a focus on estimating the total amount of influence from websites.

We collected data from human subjects answering truth questions in teams of four. After individually answering a question, they then collaborated to agree on a final answer through a chat system. The participants were periodically asked to assess their appraisals of each other. We built a machine learning-based model using text content, the time of messages, and individual task performance to estimate the collective influence matrix. We sought to find underlying factors that contribute to the accorded influence. We proposed a dynamical cognitive model, a linear model using convex optimization, and a neural network model alongside baselines from dynamical models and sociology literature to test our hypotheses. From these findings, we concluded that task performance and higher values of confidence were the two most salient factors in determining the amount of influence one receives in collaborative group settings. We believe this study on estimating underlying influence systems in a collaborative environment will spur the establishment of connections with a variety of fields and advance an interdisciplinary understanding of the design of social experiments.


\section*{Supporting information}

\paragraph*{S1 Appendix.}
\label{S1_Appendix}
{\bf Proofs of lemmas from methods.} Additional remarks on cognitive dynamical models and proofs for all lemmas introduced in main paper.

\paragraph*{S2 Appendix.}
\label{S2_Appendix}
{\bf Additional figures and results.} Simulations of equilibria of cognitive dynamical models, figures of pearson correlation of metrics defined in Definitions, and tables of additional details of results for Fig.~\ref{fig_granger-causality-result}, Fig.~\ref{fig_mse-models-with-features} and Fig.~\ref{fig_comparison-of-all-models}.





\section*{Acknowledgments}
We thank Dr. Fabio Fagnani and Dr. Giacomo Como for their invaluable discussions with authors regarding the dynamical cognitive models. 


\nolinenumbers

\begin{huge}
\centerline{S1 Appendix: Proof of Lemmas}
\end{huge}

\section{Methods}
In this section, we provide remarks on the cognitive dynamical model and rigorous proofs for every lemma introduced in the main paper.

\subsection{Auxiliary remarks for the cognitive dynamical model}\hfil\\
As defined in the main text, the Differentiation, Reversion, Perceived expertise (DRP) model is written in matrix form as
    \begin{equation}
    \label{si:model:diff-rev-skewed}
        \hat M^{(t+1)} = (1-\tau)\hat M^{(t)} +  \tau\Big( \hat y^{(t)} \hat y^{(t)\top} +  \frac{1}{n} \big(\ones[n]-\hat y^{(t)}\big) \ones[n]^\top \Big),
    \end{equation}
where $\hat M_d^{(t)} = [\hat M_{11}^{(t)},\dots, \hat M_{nn}^{(t)}]^\top \in[0,1]^{n}$ denotes the vector of self-influence weights of the influence matrix at time $t\geq 1$ and $\hat y^{(t)} = (y^\top \hat M_d^{(t)})^{-1} \diag(y) \hat M_d^{(t)} \in\Delta_n$ denotes the normalized perceived expertise.

For the DRP model, the dynamics of the self-influence matrix weights is closed from the dynamics of the interpersonal influence weights, and can be written as only a function of $\hat M_d^{(t)}$,
\begin{equation}
\label{si:model:drp-diagonal}
    \hat M_{d}^{(t+1)} = (1-\tau) \hat M_{d}^{(t)} 
        + \tau \Big( \diag(\hat{y}^{(t)}) \hat{y}^{(t)} + \diag(\ones[n]-\hat{y}^{(t)}) \frac{1}{n}\ones[n] \Big).
\end{equation}
Then the dynamics of the interpersonal influence evolve as a function of the self-influence weights and the initial condition $\hat M^{(1)}$. This cascading effect illustrates how confidence drives influence. 

\subsection{Proof of Lemma 1}\hfil\\
Per our assumption that there exists at least one $i$ such that $\hat M_{ii}^{(1)}>0$ and $y_{i}>0$, then $y^\top \hat M_d^{(0)}>0$. Additionally, the DRP model guarantees that $\hat M_{ij}^{(t)} \geq (1-\tau)^{(t-1)}\hat M_{ij}^{(1)} \geq 0$ for all $i,j$ and finite time $t\geq 2$. 

Next we show that the DRP dynamics preserves row-stochasticity. By definition, the sum of the perceived expertise is $\ones[n]^\top \hat y^{(t)} = \big(y^\top \hat M_d^{(t)}\big)^{-1} y^\top \hat M_d^{(t)} = 1$. Then, we have
\begin{align*}
    \hat M^{(t+1)}\ones[n] &= (1-\tau)\hat M^{(t)}\ones[n] +  \tau\Big( \hat y^{(t)} \hat y^{(t)\top}\ones[n] +  \frac{1}{n} \big(\ones[n]-\hat y^{(t)}\big) \ones[n]^\top\ones[n] \Big)\\
    &= (1-\tau)\ones[n] +  \tau\Big( \hat y^{(t)} +  \big(\ones[n]-\hat y^{(t)}\big) \Big)\\
    &= (1-\tau)\ones[n] + \tau \ones[n] = \ones[n].
\end{align*}
Therefore, $\hat M^{(t)}$ remains row-stochastic and well-posed for finite time $t\geq 1$.
\qed

\subsection{Proof of Lemma 2}\hfil\\
Since the dynamics of $\hat M^{(t+1)}$ can be described as a function of $M_{d}^{(t)}$, then it is sufficient to prove convergence of the self-influence weights given by the dynamics~\eqrefreg{si:model:drp-diagonal}. Note that $\hat y^{(t)} = (y^\top \hat M_{d}^{(t)})^{-1} \diag(y)\hat M_{d}^{(t)} = (c y^\top \hat M_{d}^{(t)})^{-1} \diag(c y)\hat M_{d}^{(t)}$ for any $c>0$. Then per our assumption that $y=c\ones[n]$, we can assume $y = \ones[n]$ without loss of generality and $\hat{y}^{(t)} = \big(\ones[n]^\top \hat{M}_d^{(t)}\big)^{-1} \hat{M}_d^{(t)}$. For our proof, first, we show that all trajectories satisfying our initial condition assumptions reach the forward-invariant set $\Omega = \setdef{\hat{M}_d^{(t)}\in[0,1]^{n}}{\ones[n]^\top \hat{M}_d^{(t)} \geq 1}$. Second, we define a function $\map{V}{[0,1]^{n}}{\real}$ that we prove is a Lyapunov function for $\hat{M}_{d}^{(t)} \in \Omega$. 

From the dynamics, all influence weight values become strictly positive for $t\geq2$, regardless of the initial condition. If $\hat{M}_{ij}^{(1)}=0$, then $\hat{M}_{ij}^{(2)} =\frac{\tau}{n}$ and if $\hat{M}_{ij}^{(1)}>0$, then $\hat{M}_{ij}^{(2)} >(1-\tau)\hat{M}_{ij}^{(1)}$.

Consider $ \hat{M}_{d}^{(t)} \notin \Omega$ where $\ones[n]^\top \hat{M}_{d}^{(t)} < 1$, then 
\begin{align*}
    \ones[n]^\top \hat{M}_{d}^{(t+1)} &= (1-\tau) \ones[n] \hat{M}_{d}^{(t)} + \tau \hat{y}^{(t)\top} \hat{y}^{(t)} + \tau - \frac{\tau}{n} 
    > \ones[n]^\top \hat{M}_{d}^{(t)} -\tau + \frac{\tau}{n} + \tau -\frac{\tau}{n} 
    = \ones[n]^\top \hat{M}_{d}^{(t)}.
\end{align*}
The inequality from the last line follows from the fact that $\hat{y}^{(t)\top} \hat{y}^{(t)} \geq \tfrac{1}{n}$, which can be found by formulating the minimization problem of $\hat{y}^{(t)\top} \hat{y}^{(t)}$ as a constrained convex optimization problem and applying the KKT conditions. Therefore if $ \hat{M}_{d}^{(t)} \notin \Omega$ and $\ones[n]^\top \hat{M}_{d}^{(t+1)} > \ones[n]^\top \hat{M}_{d}^{(t)}$, there exists some finite time $T>t$ such that $\hat{M}_{d}^{(T)} \in \Omega$. Next, consider  $\hat{M}_{d}^{(t)}\in\Omega$ implies
\begin{align*}
    \ones[n]^\top \hat{M}_{d}^{(t+1)} \geq (1-\tau)\ones[n]^\top \hat{M}_{d}^{(t)} - \tau + \frac{\tau}{n} + \tau - \frac{\tau}{n} \geq 1.
\end{align*}
Since the dynamics also preserves row-stochasticity of the influence network by Lemma~1, then $\ones[n]^\top \hat{M}_{d}^{(t)} \leq n$ and we have shown that $\Omega$ is a compact forward-invariant set, where trajectories that enter $\Omega$ remain in $\Omega$.

Now we prove that all trajectories in $\Omega$ converge to the equilibrium $\hat{M}_{d}^{(*)}$. It is straightforward to verify that the equilibrium of~\eqref{si:model:drp-diagonal} is $\hat M_{d}^{(*)} = \tfrac{1}{n}\ones[n]$, which corresponds to the influence network equilibrium $\hat M^{(*)}=\frac{1}{n}\ones[n]\ones[n]^\top$. We define $V(\hat{M}_{d}^{(t)})$ as
\begin{equation*}
    V(\hat{M}_{d}^{(t)}) = \max_{i\in\{1,\dots,n\}}\Big\{ \hat{M}_{ii}^{(t)} - \frac{1}{n} \Big\},
\end{equation*}
where $V(\tfrac{1}{n}\ones[n]) = 0$ and $V(\hat{M}_{d}^{(t)}) > 0$ for $\hat{M}_{d}^{(t)} \in \Omega \setminus \tfrac{1}{n}\ones[n]$. Next we show that $V(\hat{M}_{d}^{(t+1)}) < V(\hat{M}_{d}^{(t)})$ for $\hat{M}_{d}^{(t)} \in \Omega \setminus \tfrac{1}{n}\ones[n]$. Note that for $\hat y^{(t)}\in\Delta_n$ and $\hat{M}_{d}^{(t)}\in\Omega$, then $\hat{y}_i^{(t)} \leq \hat{M}_{ii}^{(t)}$ and $\max_{i}\{\hat{M}_{ii}^{(t)}\} \geq \tfrac{1}{n}$, which are used for the following bounds.
\begin{align*}
    V(\hat{M}_{d}^{(t+1)}) - V(\hat{M}_{d}^{(t)}) &= \max_{i\in\{1,\dots,n\}}\Big\{ \hat{M}_{ii}^{(t+1)} - \frac{1}{n} \Big\} - \max_{i\in\{1,\dots,n\}}\Big\{ \hat{M}_{ii}^{(t)} - \frac{1}{n} \Big\} \\ 
    &\leq (1-\tau)\max_{i\in\{1,\dots,n\}}\Big\{ \hat{M}_{ii}^{(t)} - \frac{1}{n} \Big\} + \tau \max_{i\in\{1,\dots,n\}}\Big\{ \hat y_{i}^{(t)}\big( \hat{y}_{i}^{(t)} - \frac{1}{n}\big) \Big\} - \max_{i\in\{1,\dots,n\}}\Big\{ \hat{M}_{ii}^{(t)} - \frac{1}{n} \Big\} \\
    &\leq -\tau \max_{i\in\{1,\dots,n\}}\Big\{ \hat{M}_{ii}^{(t)} - \frac{1}{n} \Big\} + \tau \hat{M}_{ii}^{(t)} \max_{i\in\{1,\dots,n\}}\Big\{ \hat{M}_{ii}^{(t)} - \frac{1}{n} \Big\} < 0.
\end{align*}
As a result, all trajectories reach $\Omega$ in finite time and trajectories in $\Omega$ approach $\lim_{t\to\infty}\max_{i}\{ \hat{M}_{ii}^{(t)} \} = \tfrac{1}{n}$, which can only occur for $\hat{M}_d^{(*)} = \tfrac{1}{n}\ones[n]$. For $\hat{M}_d^{(t)} = \hat{M}_d^{(*)}$, the dynamics of the interpersonal influence weights simplify to a stable affine system, 
\begin{equation*}
    \hat{M}_{ij}^{(t+1)} = (1-\tau)\hat{M}_{ij}^{(t)} + \tau \Big( \hat{y}_i^{(t)} \hat{y}_j^{(t)} + \big(1-\hat{y}_i^{(t)}\big)\frac{1}{n} \Big).
\end{equation*}
Therefore, for any $\hat M^{(1)}\in[0,1]^{n\times n}$ with at least one strictly positive diagonal entry, then $\lim_{t\to\infty}\hat M^{(*)}=\tfrac{1}{n}\ones[n]^\top \ones[n]$.
\qed

\subsection{Proof of Lemma 3}\hfil\\
With respect to variables $\matrixSym{W}_k$ for any $k= 1 \text{ to } K$ and $B$, the loss function $\sum\limits_{m=1}^{N}\sum\limits_{t=1}^T \Big\lVert \sum\limits_{k=1}^{K} \matrixSym{X}^{(m, t)}_k \matrixSym{W}_k^T
+ \matrixSym{B} - \matrixSym{M}^{(m, t)}\Big\rVert_F^2$ is a summation of multiple squared Frobenius norms. The regularization term $\sum\limits_{k=1}^{K} \lVert\matrixSym{W}_k\rVert_{1, 1}
+ \lVert\matrixSym{B}\rVert_{1, 1}$ is also a summation of $l1-$norms. All of which are convex functions~\cite{boyd2004convex}. Both constraints $\sum\limits_{m=1}^{N}\sum\limits_{t=1}^T
\sum\limits_{k=1}^{K} \matrixSym{X}^{(m, t)}_k \matrixSym{W}_k^T + \matrixSym{B} \geq 0$ and $\ones[n]^T \big(\sum\limits_{m=1}^{N}\sum\limits_{t=1}^T
\sum\limits_{k=1}^{K} \matrixSym{X}^{(m, t)}_k \matrixSym{W}_k^T + \matrixSym{B}\big) = \ones[n]^T$ are linear combination of variables and hence affine functions. To this end, all of the inequality constraints are convex, and all equality constraints are affine. Therefore, the problem is convex, it has a globally optimal solution, and we can solve this equation with a convex optimization solver, CVXPY~\cite{diamond2016cvxpy}.
\qed

\subsection{Proof of Lemma 4}\hfil\\
The objective function using cross-entropy can be derived, step by step, as follows,

\begin{align*}
\begin{split}
    \text{loss} &= \sum\limits_{m=1}^{N}\sum\limits_{t=1}^{T} \sum\limits_{j=1}^{n} H(M_{j, .}^{(m, t)}, \hat M_{j, .}^{(m, t)})\\
    &= \sum\limits_{m=1}^{N}\sum\limits_{t=1}^{T} \sum\limits_{j=1}^{n} -\sum\limits_{k=1}^{n} M_{j, k}^{(m, t)}\log \hat M_{j, k}^{(m, t)}\\
    &= -\sum\limits_{m=1}^{N}\sum\limits_{t=1}^{T} \sum\limits_{j=1}^{n} \sum\limits_{k=1}^{n} M_{j, k}^{(m, t)}\log \sigma(O_{j, k}^{(m, t)})\\
    &= -\sum\limits_{m=1}^{N}\sum\limits_{t=1}^{T} \sum\limits_{j=1}^{n} \sum\limits_{k=1}^{n} M_{j, k}^{(m, t)}\log \frac{\exp(O_{j, k}^{(m, t)})}{\sum\limits_{l=1}^{n} \exp(O_{j, l}^{(m, t)})}\\
    &= -\sum\limits_{m=1}^{N}\sum\limits_{t=1}^{T} \sum\limits_{j=1}^{n} \sum\limits_{k=1}^{n} M_{j, k}^{(m, t)}\bigg( O_{j, k}^{(m, t)} - \log\sum\limits_{l=1}^{n} \exp(O_{j, l}^{(m, t)}) \bigg)\\
    &= -\sum\limits_{m=1}^{N}\sum\limits_{t=1}^{T} \sum\limits_{j=1}^{n} \sum\limits_{k=1}^{n} M_{j, k}^{(m, t)}\bigg(X_{j, k}^{(m, t)} W_{k, j} + b_j - \log\sum\limits_{l=1}^{n} \exp( X_{j, k}^{(m, t)} W_{k, j} + b_j ) \bigg).
\end{split}
\end{align*}

\noindent
Therefore, it results to
\begin{align*}
\begin{split}
    \text{objective} &= \text{loss} + \lambda\bigg(\|\matrixSym{W}\|_1 + \|b\|_1\bigg)\\
    &= -\sum\limits_{m=1}^{N}\sum\limits_{t=1}^{T} \sum\limits_{j=1}^{n} \sum\limits_{k=1}^{n} M_{j, k}^{(m, t)}\bigg(X_{j, k}^{(m, t)} W_{k, j} + b_j - \log\sum\limits_{l=1}^{n} \exp( X_{j, k}^{(m, t)} W_{k, j} + b_j ) \bigg) + \lambda\bigg(\|\matrixSym{W}\|_1 + \|b\|_1\bigg)
\end{split}
\end{align*}

\noindent
The final equation for the objective function is the same as~\eqref{eq:generalized_vectorized_convex_optimization_model} from the main paper.
\qed

\subsection{Proof of Lemma 5}\hfil\\

With respect to variables $W, b$, the problem is a summation of an affine function ($-X_{j, k}^{(m, t)} W_{k, j}$), log-sum-exp term ($\log\sum\limits_{l=1}^{n} \exp( X_{j, k}^{(m, t)} W_{k, j} + b_j )$) and two $l1-$norms regularization terms $\big(-\lambda\left(||W||_1 + ||b||_1\right)\big)$. It has been proved mathematically that the expression $log\sum\exp(r)$ for any real $r$ is convex in $R^n$ ~\cite{boyd2004convex}. In here, $r$ is an affine function $-X_{j, k}^{(m, t)} W_{k, j}$ with respect to variables $W, b$. Hence, the problem is a summation of an affine and two convex terms which preserves convexity and produces a convex function~\cite{boyd2004convex}. Therefore, the minimization problem is convex, it has a globally optimal solution, and we can solve this equation with the aforementioned convex optimization solver.
\qed
\begin{huge}
\centerline{S2 Appendix: Additional Results }
\end{huge}

\section{Results}

\begin{figure}[!ht]
    \centering
    \begin{subfigure}[t]{0.45\linewidth}
    \centering
        \includegraphics[width=\linewidth]{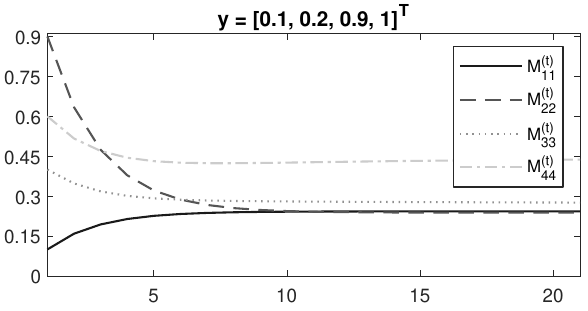}
    \end{subfigure}~\begin{subfigure}[t]{0.45\linewidth}
    \centering
        \includegraphics[width=\linewidth]{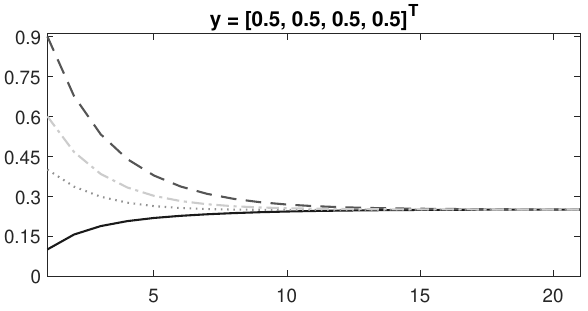}
        
    \end{subfigure}

    \caption{For a team of 4 individuals, the plots depict the evolution of the self-influence matrix weights $\hat M_{ii}$ for all $i\in\{1,\dots,4\}$ obeying the DRP model (S1 Appendix Eq.~\eqrefreg{si:model:diff-rev-skewed}) with constant expertise $y$ as specified and time scale $\tau=0.4$. In general, the self-influence overestimates low-performers and underestimates high-performers. However, when all individuals have the same expertise level, then $\hat M^{(t)}$ converges to $\frac{1}{n}\ones[n]\ones[n]^\top$.}
    \label{fig:dynamicModels-evolution}
\end{figure}

Here we describe the simulation results for equilibrium of the proposed dynamical model. Fig.~\ref{fig:dynamicModels-evolution} illustrates how the self-influence weights of the DRP model (S1 Appendix~\eqrefreg{si:model:diff-rev-skewed}) with constant expertise $y$ and time scale $\tau=0.4$ evolve over time. The interpersonal influence weights are not plotted, since they are a function of $\hat M^{(1)}$, $\hat M_{d}^{(t)}$, and $y$.

\begin{figure}[!ht]
    \centering
    \includegraphics[width=0.5\linewidth]{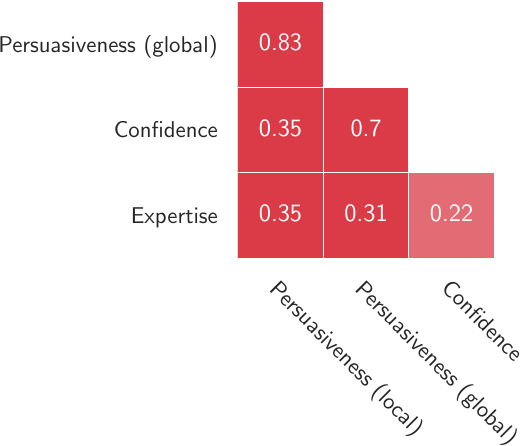}
    \caption{Pearson correlation ($r$-value) of metrics on influence and expertise after answering all 45 questions with statistical significance of $p\text{-value} < 0.018$. All metrics are formally presented in the main paper. The threshold for $p$-value is chosen using Benjamini-Hochberg (BH) procedure with False Discovery Rate of 5\%~\cite{benjamini1995controlling}. Persuasiveness for every subject in the final reported influence matrix vs. the expertise by the same subject has the coefficients $r$-value = 0.35 and $p$-value < 2e-4. This result shows that expertise is statistically correlated with the amount of appraisal one would receive by others in a team; this phenomenon is consistent with research on Transactive Memory System (TMS)~\cite{lewis2005transactive, lewis2003measuring, lewis2004knowledge}. Surprisingly, this also shows in about two hours with indirect performance exposure, team members are still able to uncover each other's expertise. Moreover, confidence and appraise by others are also statistically correlated which is aligned with the research in confidence heuristics~\cite{thomas1995confidence, price2004intuitive}.}
    \label{fig:correlation_results}
\end{figure}

Fig.~\ref{fig:correlation_results} demonstrates the Pearson correlations of every pair of metrics defined in Definitions. In this figure, all correlations are statistically significant as their corresponding $p$-values have been corrected using the Benjamini-Hochberg (BH) procedure~\cite{benjamini1995controlling} with False Discovery Rate of 5\% has the required $p\text{-value} < 0.018$ as the statistical significance threshold. Recall that the global perception (eigenvector) is given by
\begin{equation} \label{eq:global_persuasiveness}
    \big(\vleft(\matrixSym{C}^{(t)})\big)_i,
\end{equation}
\noindent where $\matrixSym{C}^{(t)} = \diag\big((\matrixSym{M}^{(t)} - \matrixSym{D}^{(t)})\ones[n]\big)^{-1}\big(\matrixSym{M}^{(t)} - \matrixSym{D}^{(t)}\big)$ shows relative interpersonal influence matrix at round $t$ where $\matrixSym{D}^{(t)}$ is a diagonal matrix with the diagonal entries of influence matrix $\matrixSym{M}^{(t)}$ at round $t$.
Note that global persuasiveness (\eqref{eq:global_persuasiveness}) is the stationary probability of the influence matrix after removing its main diagonal. This result shows that expertise is statistically correlated with the amount of appraisal one would receive by others in a team.

For more details on the results shown in Fig.~\ref{fig_mse-models-with-features} from the main paper, we provide Table~\ref{tbl:mse_errors}. Also for more details on the results shown in Fig.~\ref{fig_comparison-of-all-models} from the main paper, we provide Table~\ref{tbl:mse_errors_previous_influence} and Table~\ref{tbl:kl_errors_previous_influence} to show MSE and KL divergence, respectively. All models use previous influence matrix and expertise. Similarly, neural network-based model surpasses all baselines and provides statistically and significantly lower error rate.

\begin{table}[!ht]
\centering
\begin{adjustbox}{max width=\textwidth}
\begin{tabular}{|c|c|c|c|c|c|c|c|c|c|c|}
\hline
\diagbox{Features}{Models} &
\begin{tabular}[c]{@{}c@{}}Random\\ baseline\end{tabular} & 
\begin{tabular}[c]{@{}c@{}}First\\ baseline\end{tabular} &
\begin{tabular}[c]{@{}c@{}}SBT\\ baseline\end{tabular} &
\begin{tabular}[c]{@{}c@{}}Uniform\\ baseline\end{tabular} & \begin{tabular}[c]{@{}c@{}}Average\\ baseline\end{tabular} & 
\begin{tabular}[c]{@{}c@{}}Reflected\\ appraisal\\ model\end{tabular} &
\begin{tabular}[c]{@{}c@{}}Constant\\ baseline\end{tabular} &
\begin{tabular}[c]{@{}c@{}}Linear\\ model\end{tabular} &
\begin{tabular}[c]{@{}c@{}}Cognitive\\ DRP\\-based\\ model\end{tabular} &  \begin{tabular}[c]{@{}c@{}}Neural\\ network\\-based\\ model\end{tabular} \\ \hline
previous influence matrix &
\begin{tabular}[c]{@{}c@{}}0.0305\\ (0.00007)\end{tabular} & \begin{tabular}[c]{@{}c@{}}0.0176\\ (0.00008)\end{tabular} & \begin{tabular}[c]{@{}c@{}}0.0172\\ (0.00007)\end{tabular} &
\begin{tabular}[c]{@{}c@{}}0.0110\\ (0.00004)\end{tabular} & \begin{tabular}[c]{@{}c@{}}0.0106\\ (0.00005)\end{tabular} &
N/A &
\begin{tabular}[c]{@{}c@{}}0.0073\\ (0.00007)\end{tabular} &
\begin{tabular}[c]{@{}c@{}}0.0061\\ (0.00005)\end{tabular} &
N/A &
\textbf{\begin{tabular}[c]{@{}c@{}}0.0059\\ (0.00003)\end{tabular}}
\\ \hline
\begin{tabular}[c]{@{}c@{}}previous influence matrix,\\ expertise\end{tabular} & \begin{tabular}[c]{@{}c@{}}0.0305\\ (0.00007)\end{tabular} & \begin{tabular}[c]{@{}c@{}}0.0176\\ (0.00008)\end{tabular} & \begin{tabular}[c]{@{}c@{}}0.0172\\ (0.00007)\end{tabular} &
\begin{tabular}[c]{@{}c@{}}0.0110\\ (0.00004)\end{tabular} & \begin{tabular}[c]{@{}c@{}}0.0106\\ (0.00005)\end{tabular} &  \begin{tabular}[c]{@{}c@{}}0.0073\\ (0.00007)\end{tabular} &
\begin{tabular}[c]{@{}c@{}}0.0073\\ (0.00007)\end{tabular} & \begin{tabular}[c]{@{}c@{}}0.0059\\ (0.00005)\end{tabular} & \begin{tabular}[c]{@{}c@{}}0.0058\\ (0.00005)\end{tabular} & \textbf{\begin{tabular}[c]{@{}c@{}}0.0058\\ (0.00003)\end{tabular}} \\ \hline
\end{tabular}
\end{adjustbox}
\caption{MSE of estimated influence matrix from the ground truth in the test set of influence matrices reported by individuals. The number in parenthesis shows the standard error in 1000 bootstrap on test set. This table also depicts the proposed learning models (neural network-based and linear) outperform baselines. Surprisingly, it also shows the reflected appraisal does not surpass the baseline of considering previous influence matrix to be unchanged (constant baseline). Also interestingly, this table shows that the cognitive based on Differentiation, Reversion, Perceived (DRP) model~\eqref{si:model:diff-rev-skewed} from S1 Appendix works competitively with the learning models showing the power behind this social phenomenon~\cite{friedkin1990social}.}
\label{tbl:mse_errors_previous_influence}
\end{table}

\begin{table}[!ht]
\centering
\begin{adjustbox}{max width=\textwidth}
\begin{tabular}{|c|c|c|c|c|c|c|c|c|c|c|}
\hline
\diagbox{Features}{Models}&
\begin{tabular}[c]{@{}c@{}}Random\\ baseline\end{tabular} & \begin{tabular}[c]{@{}c@{}}First\\ baseline\end{tabular} &
\begin{tabular}[c]{@{}c@{}}SBT\\ baseline\end{tabular} &
\begin{tabular}[c]{@{}c@{}}Uniform\\ baseline\end{tabular} & \begin{tabular}[c]{@{}c@{}}Average\\ baseline\end{tabular} & 
\begin{tabular}[c]{@{}c@{}}Reflected\\ appraisal\\ model\end{tabular} &
\begin{tabular}[c]{@{}c@{}}Constant\\ baseline\end{tabular} & \begin{tabular}[c]{@{}c@{}}Linear\\ model\end{tabular} &
\begin{tabular}[c]{@{}c@{}}Cognitive\\ DRP\\-based\\ model\end{tabular} & \begin{tabular}[c]{@{}c@{}}Neural\\ network\\ -based\\ model\end{tabular} \\ \hline
previous influence matrix &
\begin{tabular}[c]{@{}c@{}}0.3431\\ (0.0008)\end{tabular} & \begin{tabular}[c]{@{}c@{}}0.1908\\ (0.0009)\end{tabular} & \begin{tabular}[c]{@{}c@{}}0.1540\\ (0.0008)\end{tabular} &
\begin{tabular}[c]{@{}c@{}}0.0818\\ (0.0003)\end{tabular} & \begin{tabular}[c]{@{}c@{}}0.0709\\ (0.0006)\end{tabular} &
N/A &
\begin{tabular}[c]{@{}c@{}}0.0707\\ (0.0006)\end{tabular} &
\begin{tabular}[c]{@{}c@{}}0.0494\\ (0.0003)\end{tabular} &
N/A &
\textbf{\begin{tabular}[c]{@{}c@{}}0.0479\\ (0.0003)\end{tabular}}
\\ \hline
\begin{tabular}[c]{@{}c@{}}previous influence matrix,\\ expertise\end{tabular} & \begin{tabular}[c]{@{}c@{}}0.3431\\(0.0008)\end{tabular} & \begin{tabular}[c]{@{}c@{}}0.1908\\ (0.0009)\end{tabular} & \begin{tabular}[c]{@{}c@{}}0.1540\\ (0.0008)\end{tabular} &
\begin{tabular}[c]{@{}c@{}}0.0818\\ (0.0003)\end{tabular} & \begin{tabular}[c]{@{}c@{}}0.0709\\ (0.0006)\end{tabular} &  \begin{tabular}[c]{@{}c@{}}0.0709\\ (0.0006)\end{tabular} &
\begin{tabular}[c]{@{}c@{}}0.0707\\ (0.0006)\end{tabular} & \begin{tabular}[c]{@{}c@{}}0.0482\\ (0.0003)\end{tabular} & \begin{tabular}[c]{@{}c@{}}0.0478\\ (0.0003)\end{tabular} & \textbf{\begin{tabular}[c]{@{}c@{}}0.0459\\ (0.0003)\end{tabular}} \\ \hline
\end{tabular}
\end{adjustbox}
\caption{KL divergence of estimated influence matrix from the ground truth in the test set of influence matrices reported by individuals. The number in parenthesis shows the standard error in 1000 bootstrap on test set. Other designations in the table are the same as Table~\ref{tbl:mse_errors_previous_influence}.}
\label{tbl:kl_errors_previous_influence}
\end{table}
\clearpage

\begin{table}[!ht]
\centering
\begin{adjustbox}{max width=\textwidth}
\begin{tabular}{|l|c|c|c|c|c|c|}
\hline
\diagbox{Features}{Models} & \begin{tabular}[c]{@{}c@{}}Random\\ baseline\end{tabular} & \begin{tabular}[c]{@{}c@{}}First\\ baseline\end{tabular} & \begin{tabular}[c]{@{}c@{}}Uniform\\ baseline\end{tabular} & \begin{tabular}[c]{@{}c@{}}Average\\ baseline\end{tabular} & \begin{tabular}[c]{@{}c@{}}Linear\\ model\end{tabular} & \begin{tabular}[c]{@{}c@{}} Neural Network-based\\ model\end{tabular} \\ \hline
expertise & \begin{tabular}[c]{@{}c@{}}0.0305\\ (0.00007)\end{tabular} & \begin{tabular}[c]{@{}c@{}}0.0176\\ (0.00008)\end{tabular} & \begin{tabular}[c]{@{}c@{}}0.0110\\ (0.00005)\end{tabular} & \begin{tabular}[c]{@{}c@{}}0.0106\\ (0.00004)\end{tabular} & \begin{tabular}[c]{@{}c@{}}0.0106\\ (0.00005)\end{tabular} & \textbf{\begin{tabular}[c]{@{}c@{}}0.0102\\ (0.00004)\end{tabular}} \\ \hline
\begin{tabular}[c]{@{}l@{}}expertise,\\ first influence matrix\end{tabular} & \begin{tabular}[c]{@{}c@{}}0.0305\\ (0.00007)\end{tabular} & \begin{tabular}[c]{@{}c@{}}0.0176\\ (0.00008)\end{tabular} & \begin{tabular}[c]{@{}c@{}}0.0110\\ (0.00005)\end{tabular} & \begin{tabular}[c]{@{}c@{}}0.0106\\ (0.00004)\end{tabular} & \begin{tabular}[c]{@{}c@{}}0.0093\\ (0.00005)\end{tabular} & \textbf{\begin{tabular}[c]{@{}c@{}}0.0079\\ (0.00003)\end{tabular}} \\ \hline
\begin{tabular}[c]{@{}l@{}}expertise,\\ first influence matrix,\\ response network\end{tabular} & \begin{tabular}[c]{@{}c@{}}0.0305\\ (0.00007)\end{tabular} & \begin{tabular}[c]{@{}c@{}}0.0176\\ (0.00008)\end{tabular} & \begin{tabular}[c]{@{}c@{}}0.0110\\ (0.00005)\end{tabular} & \begin{tabular}[c]{@{}c@{}}0.0106\\ (0.00004)\end{tabular} & \begin{tabular}[c]{@{}c@{}}0.0093\\ (0.00005)\end{tabular} & \textbf{\begin{tabular}[c]{@{}c@{}}0.0072\\ (0.00003)\end{tabular}} \\ \hline
\begin{tabular}[c]{@{}l@{}}expertise,\\ first influence matrix,\\ response network,\\ sentiment network,\\ emotion network\end{tabular} & \begin{tabular}[c]{@{}c@{}}0.0305\\ (0.00007)\end{tabular} & \begin{tabular}[c]{@{}c@{}}0.0176\\ (0.00008)\end{tabular} & \begin{tabular}[c]{@{}c@{}}0.0110\\ (0.00005)\end{tabular} & \begin{tabular}[c]{@{}c@{}}0.0106\\ (0.00004)\end{tabular} & \begin{tabular}[c]{@{}c@{}}0.0093\\ (0.00004)\end{tabular} & \textbf{\begin{tabular}[c]{@{}c@{}}0.0070\\ (0.00003)\end{tabular}} \\ \hline
\begin{tabular}[c]{@{}l@{}}expertise,\\ first influence matrix,\\ response network,\\ sentiment network,\\ emotion network,\\ content embedding\end{tabular} & \begin{tabular}[c]{@{}c@{}}0.0305\\ (0.00007)\end{tabular} & \begin{tabular}[c]{@{}c@{}}0.0176\\ (0.00008)\end{tabular} & \begin{tabular}[c]{@{}c@{}}0.0110\\ (0.00005)\end{tabular} & \begin{tabular}[c]{@{}c@{}}0.0106\\ (0.00004)\end{tabular} & \begin{tabular}[c]{@{}c@{}}0.0090\\ (0.0004)\end{tabular} & \textbf{\begin{tabular}[c]{@{}c@{}}0.0069\\ (0.00003)\end{tabular}} \\ \hline
\begin{tabular}[c]{@{}l@{}}expertise,\\ first influence matrix,\\ response network,\\ sentiment network,\\ emotion network,\\ content embedding,\\ average of previous\\ influence matrices\end{tabular} & \begin{tabular}[c]{@{}c@{}}0.0305\\ (0.00007)\end{tabular} & \begin{tabular}[c]{@{}c@{}}0.0176\\ (0.00008)\end{tabular} & \begin{tabular}[c]{@{}c@{}}0.0110\\ (0.00005)\end{tabular} & \begin{tabular}[c]{@{}c@{}}0.0106\\ (0.00004)\end{tabular} & \begin{tabular}[c]{@{}c@{}}0.0076\\ (0.00005)\end{tabular} & \textbf{\begin{tabular}[c]{@{}c@{}}0.0066\\ (0.00003)\end{tabular}} \\ \hline
\end{tabular}
\end{adjustbox}
\caption{\textbf{Improvement in machine learning models by adding more features:} Average of Mean Square Error (MSE) of estimated influence matrix from the ground truth in the test set of influence matrices reported by individuals. The number in parenthesis shows the standard error in 1000 bootstrap on test error.}
\label{tbl:mse_errors}
\vsa
\end{table}

\subsubsection*{Correlation study of interpersonal influence}
To study the relationship between expertise and persuasiveness, we compute the correlation between the amount of persuasiveness with expertise after answering all questions. We use the final influence matrix reported by every team and their expertise at the end of the experiment. Note that, this data satisfies correlation, regression, and causation studies' requirements since the amount of influence one reports for themselves, other teammates report about that person, and their individual correct answer rate are independent and identically distributed.
We report empirical evidence for all the aforementioned hypotheses. Pearson correlation results show that there is a statistically positive correlation between expertise and persuasiveness ($r\text{-value} = 0.35$ and $p\text{-value} < 2e-4$) (see S2 Appendix  Fig.~\ref{fig:correlation_results}). This is the statistical evidence for Hypothesis~\ref{hp:hypothesis1} from the main text. This phenomenon was predicted by cognitive science and evolutionary anthropology studies depicting individuals naturally engage in selective social learning~\cite{almaatouq2020adaptive, wisdom2013social, boyd2011cultural}. This finding is also an empirical result for the seminal research in the theory of TMS~\cite{lewis2005transactive, lewis2003measuring, lewis2004knowledge}. We provide an empirical evidence for Hypothesis~\ref{hp:hypothesis2} from the main text via predicting mean reversion (\eqref{eq:reversion} in the main text) using expertise in the following subsection. Finally, we also find that there is a statistical positive correlation between persuasiveness and confidence ($r\text{-value} = 0.22$ and $p\text{-value} < 2e-2$). This is the statistical evidence for Hypothesis~\ref{hp:hypothesis3} from the main text. This result is consistent with reported empirical findings~\cite{almaatouq2020adaptive, becker2017network, madirolas2015improving}. Becker \emph{et al.}~\cite{becker2017network} and Almaatouq \emph{et al.}~\cite{almaatouq2020adaptive} both measure confidence as proportionate to the change in the individual answer before and after the discussion (so-called "weight on self") and present evidence for its correlation with one's popularity. This is also aligned with past research in confidence heuristics~\cite{thomas1995confidence, pulford2018persuasive, mercier2012two, hertwig2012tapping, price2004intuitive}. These two significant correlation were replicated in a different empirical experiment on wisdom of crowds where the network structure was dynamic~\cite{almaatouq2020adaptive}. However, in this experiment, the result is surprising since individuals communicate only through chat and nonetheless the expertise and the self-confidence impacted people's judgment stronger than any cognitive biases. Also, our experiments show the type of definition for persuasiveness does not change the sign of this relationship, nor its statistical significance. Thus, due to the large correlation between global and local perceptions of persuasiveness, the similarity of their distributions, and more straightforward local definition, we use its local definition in the rest of the paper.

\subsubsection*{Additional results on causality of interpersonal influence}
To study the order of effects given in Fig.~\ref{fig_granger-causality-result} in the main text, we use the proportion of statistically significant Granger causality associated with different lags. Every lag depicts five questions, as we inquire about their influence matrix every five questions. In this regard, we test the causality once using lag=1 and then by using lag=2. We analyze the proportion of statistical significance only with lag=1 and the rest that needs at least two previous time data points (lag=2). Thus, we sort the effects based on the descending order of the proportion of lag=1 as an estimate for the underlying order. From the effects in Fig.~\ref{fig_granger-causality-result} from the main text, Expertise $\longrightarrow$ Confidence seems to be the fastest (as it has the most number of statistical significance effects with lag=1 compared to the rest of the effects). Afterward, Confidence $\longrightarrow$ Persuasiveness and Expertise $\longrightarrow$ Persuasiveness both come very close to each other. Lastly, Persuasiveness $\longrightarrow$ Confidence seems the slowest among all. in other words, the order shows that the expertise of individuals quickly impacts their confidence. Both their confidence and their expertise then lead to persuasiveness. However, it seems that confidence has a slightly faster effect. This is perfectly aligned with the past research in confidence heuristic~\cite{thomas1995confidence} and means that people have an immediate effect by social power and confidence, indeed faster than expertise. However, still, after some time (two lags in this experiment) expertise as a heuristic would be more prominent in the amount of persuasiveness (as 70\% of individuals have this causal relationship after maximum two lags). In the end, persuasiveness can also lead to confidence; however, although it happens very often, it happens slower than the aforementioned effects. The delay of the persuasiveness affecting confidences could be because the platform shares individuals' answers with everyone but does not disclose their appraisals of each other. Therefore individuals must learn their confidence from persuasiveness through multiple discussions.



\end{document}